\documentclass[12pt,a4paper]{article}   
\usepackage{soul}

\usepackage[utf8]{inputenc}
\usepackage[english]{babel}

\usepackage{tabularx} 

\usepackage[theorems, global-numbering, colorlinks]{default}

\usepackage[backend=biber,style=alphabetic]{biblatex}
\usepackage[normalem]{ulem}

\usepackage[left=2cm,right=2cm,top=2.5cm,bottom=2cm,headheight=15pt]{geometry}

\usepackage{dsfont}

\usepackage{orcidlink}

\DeclareMathOperator{\Corr}{Corr}

\newcommand{\e}{\mathrm e}

\title{International Financial Markets Through 150 Years: Evaluating Stylized Facts}

\author{Sara A. Safari\footnote{School of Engineering, Forschungsschwerpunkt Finance, Risk Management and Econometrics, ZHAW.}
\footnote{Department of Mathematical Modeling and Machine Learning ($DM^3L$), Universität Zürich. E-Mail: \url{sara.aliqolizadehsafari@uzh.ch}}\,\orcidlink{0000-0003-0376-7429}\and Maximilian Janisch\footnote{Institut für Mathematik, Universität Zürich. E-Mail: \url{maximilian.janisch@math.uzh.ch}}\,\orcidlink{0000-0002-7018-8833}\and Thomas Lehéricy\footnote{Physik Institut, Universität Zürich. E-Mail: \url{thomas.lehericy@math.uzh.ch}}\,\orcidlink{0000-0001-7508-7233}}
\date{\today}

\begin{document}
	\maketitle
    \begin{abstract}
         In the theory of financial markets, a \emph{stylized fact} is a qualitative summary of a pattern in financial market data that is observed across multiple assets, asset classes and time horizons. In this article, we test a set of eleven stylized facts for financial market data. Our main contribution is to consider a broad range of geographical regions across Asia, continental Europe, and the US over a time period of 150 years, as well as two of the most traded cryptocurrencies, thus providing insights into the robustness and generalizability of commonly known stylized facts. 
    \end{abstract}
	\tableofcontents

    \section{Introduction}
    \label{sect:intro}
    A \emph{stylized fact} is a simplified presentation of an empirical pattern in data that captures broad tendencies. A stylized fact transcends changes in datasets, timeframes, measurement methods, and geographical locations; its value lies in its consistency across these dimensions, even though specific counterexamples may exist. 
    Notable examples include Zipf's law \cite{zipf} in linguistics, where word frequencies follow a power-law distribution, and similar power-law patterns observed in phenomena such as city populations and earthquake magnitudes \cite{PerBak_book}. Another classic example is the economic observation that \enquote{Education significantly raises lifetime income}, which holds true in general despite potential exceptions for specific individuals or career paths.

    In financial markets, stylized facts 
    were first recognized in the pioneering work of Mandelbrot \cite{mandelbrot66}, who identified scale-free behavior in commodity price fluctuations. Since then, researchers have documented numerous regularities in financial time series that transcend specific market mechanisms or participant behaviors, see e.g. \cite{Cont2001}.
    
    Stylized facts serve as powerful reference points for model evaluation, regardless of our understanding of the underlying mechanisms \cite{fernando2019, JurgOst_review}. This makes them particularly valuable in financial markets for validating mathematical models and evaluating data-augmentation techniques. Given that we typically observe only one realization of market data, stylized facts enable meaningful assessment of simulated processes when direct point-by-point comparisons are unfeasible.
    
    \cite{Cont2001}, building upon suggestions from earlier authors, proposed a set of eleven stylized facts for financial market data, understood in this context as qualitative patterns observed consistently in multiple assets, asset classes, as well as multiple regions. Although the concept of stylized facts was implicitly assumed in earlier works (\cite{andersen_boll1997, Arneodo, Bjorn94, Bouchaud_universality_EV, Bouchaud_2001, Bouchaud_stylized, Dacorogan_1993, intermittency_turbulence, Cont_Beyond, Liu97}, and many more), it was explicitly articulated in \cite{Cont2001} for the first time. These features are:
    \begin{enumerate}
        \item absence of autocorrelations of returns (see Section~\ref{sect:absence-of-autocorrelations}),
        \item slow decay of correlations (see Section~\ref{sect:slow-decay-of-autocorrelations}),
        \item intermittency (see Section~\ref{sect:intermittency}),
        \item realized volatility clustering for different metrics (see Section~\ref{sect:volatility-clustering}),
        \item cross anti-correlation of realized volatility and returns (see Section~\ref{sect:leverage-effect}),
        \item cross correlation of volume of the trade and realized volatility measures (see Section~\ref{sect:volume-volatility}),
        \item conditional heavy tails of the return distributions (see Section~\ref{sect:cond}),
        \item unconditional heavy tails of the return distributions (see Section~\ref{sect:uncond}),
        \item asymmetric tails exponent of the aforementioned distribution (left tail being heavier) (see Section~\ref{sect:gain-loss-asymmetry}),
        \item aggregational gaussianity (see Section~\ref{sect:aggregational-gaussianity}),
        \item asymmetric information flow across different time scales (see Section~\ref{sect:ts_asym}).
    \end{enumerate}

    What distinguishes \cite{Cont2001} from earlier work is the emphasis on the \enquote{universality} of these patterns, a notion from statistical mechanics. The intuition is that the aggregation of microscopic market agent behaviors often leads to similar macroscopic outcomes, even if the exact dynamics of individual agents remain unclear (see \cite{abm_doyne} for a comprehensive discussion).

    To the best of our knowledge, no single stochastic model manages to reproduce all eleven stylized facts. The celebrated Black--Scholes model, based on geometric Brownian motion, only reproduces two out of eleven. The same holds for the Ornstein--Uhlenbeck process (OU) used for interest rate modeling. Generalizations of Black--Scholes such as stochastic volatility models (e.g., the Heston model \cite{Heston1993}), jump-diffusion models, ARCH and GARCH \cite{ARCH, GARCH}, and multifractal models (e.g., MMAR \cite{mandelbrot1997multifractal}), have been introduced to capture more stylized facts like volatility clustering, heavy tails, gain/loss asymmetry, and aggregational Gaussianity. 
    Although Black--Scholes or OU capture the bulk behavior of the actual process with good precision, they exhibit serious shortcomings when used to generate the intermittent nature of markets, which is an ongoing regime change between periods of high and low volatility in various markets \cite{Grahovac_2016}. 
    Another approach, based on the observation that market prices emerge from the interactions of many participants, explores agent-based models that derive asset price models from assumptions about these interactions. 
    \cite{Katahira_spec_game} provides a model reproducing ten out of the eleven stylized facts. 

    While stylized facts provide valuable benchmarks for both theoretical modeling and empirical validation, their validity is fundamentally constrained by the evidence supporting them.    
    Empirical tests of stylized facts have been performed repeatedly in the literature, see for instance \cite{Chakraborti2009} for a review, as well as \cite{Stylized_modern} where intraday returns from October 2018 to March 2019 for the constituents of the Dow Jones Industrial were examined for evidence of Cont's facts. 
    \cite{Stylized_modern} found evidence for eight of the eleven facts, but did not find evidence for the remaining three. See also \cite{JurgOst_crypt,sherkar23}. 

    In this work, and in line with this literature, we systematically evaluate which stylized facts are supported by evidence across different markets and time periods, and to what extent. Our contribution lies in exploring a larger set of financial data: we use approximately 150 years of traditional financial market data from diverse regions, and also investigate cryptocurrencies, which have different price formation mechanisms than classical financial assets (see \cite{Stylized_BitCoin1,ZHANG2019598}).

    Our analysis provides empirical support for seven of the eleven stylized facts. We find mixed evidence for the absence of log-return autocorrelation (Section~\ref{sect:absence-of-autocorrelations}, differing between traditional markets and cryptocurrencies) and the leverage effect (Section~\ref{sect:leverage-effect}, not uniformly immediate across all assets). The results for aggregational Gaussianity (Section~\ref{sect:aggregational-gaussianity}) suggest convergence towards normality, but this may be influenced by observing fewer tail events over longer horizons. We also investigate intermittency (Section~\ref{sect:intermittency}). As standard quantitative tests are lacking, we propose specific comparisons to benchmark processes, and based on these, we find evidence supporting intermittency in the data (see Table~\ref{tab:summary_facts} for a detailed summary).

    The article is structured as follows. 
    In section \ref{sect:Methods} we discuss the methodological foundations for the hypothesis tests we perform and define the key quantities that we refer to throughout the work. 
    The data description and our code availability are discussed in section \ref{sect:data and code} for clarity and reproducibility concerns. 
    A detailed description of each stylized facts, how we choose to test them, and the results of our tests are reported in section \ref{sect:Results}. 
    We compare the traditional markets against cryptocurrencies to check for the validity of the features in these newly emerged markets. 
    The aim is to identify any discrepancy between what we observe across different time scales and regions with what has already been observed in the literature.

    \section{Methodological foundations}\label{sect:Methods}\label{sect:framework}\label{sect:volatility}
    In this section, we describe the theoretical framework for the asset prices which we use to define the statistical tests in section \ref{sect:Results}. This framework serves only to make precise some important notions such as \enquote{returns} or \enquote{volatility}. The tests themselves are model-agnostic and we make no assumptions on the specific structure of the asset returns.
    
    For a fixed asset, denote its price at time $t$ by $S_t$ for all times $t$ in some interval $I=[0,T]$ for some $T>0$. We let the $S_t$ be random variables, so that the $(S_t)_{t\in I}$ form a stochastic process. The return of the asset between two times $t,\tilde t\in I$ is $R_{t,\tilde t}\define \frac{S_{\tilde t}}{S_t}$. Assuming that $S_t>0$ for all $t\in I$, we may also consider the logarithmic returns denoted by a lowercase letter, $r_{t, \tilde t}\define\ln R_{t,\tilde t} = \ln(S_{\tilde t}) - \ln(S_t)$.

    The process $S=(S_t)_{t\in I}$ is often modeled by an Itô diffusion, i.e. we assume that there exists a standard Brownian motion $(W_t)_{t\in I}$ such that
    \begin{equation}\label{eq:Ito-diffusion}
        S_t = \int_0^t \sigma_s \,\mathrm dW_s + \int_0^t b_s\,\mathrm ds,
    \end{equation}
    for some progressively measurable stochastic processes $\sigma = (\sigma_s)_{s\in I}$ and $b = (b_s)_{s\in I}$ satisfying
    \begin{equation*}
        \int_0^t (\sigma_s^2+\lvert b_s \rvert)\,\mathrm ds < \infty
    \end{equation*}
    for every $t\in I$. We denote by $\sigma_t^2$ the \emph{instant volatility of $S$ at time $t$}. 

    When observing market data, we do not observe $\sigma$ directly but instead need to estimate the volatility. We now present different approaches to do so. Let $n\ge 2$, and fix $t_1<t_2<\dots<t_{n+1}$ a sequence of times in $I$ such that $t_{i+1}-t_i$ does not depend on $i$; we typically assume that they are chosen in such a way that $\sigma_t^2$ is constant over the interval $[t_1, t_{n+1}]$. The three estimators below provide an approximation of $\sigma_t^2$.

    \paragraph{Basic estimator}
    The \enquote{basic estimator} of the volatility is the standard sample variance estimator:
    \label{eq:basic}
	\begin{equation}
		\frac 1{n-1} \sum_{k=1}^n \left(\ln \frac{S_{k+1}}{S_k} - \frac 1 n \sum_{j=1}^n \ln\frac{S_{j+1}}{S_j}\right)^2 
        = \frac 1{n-1} \sum_{k=1}^n \left(r_{t_k, t_{k+1}}-\frac{r_{t_1, t_{n+1}}}n\right)^2 . 
	\end{equation}

    \paragraph{Parkinson estimator} Denote by $S_i^\uparrow$ the supremum of the asset price between $t_i$ and $t_{i+1}$, as well as by $S_i^\downarrow$ its infimum on this time interval:
    \begin{equation*}
        S_i^\uparrow\define\sup_{t\in[t_i, t_{i+1}]} S_t, \qquad S_i^\downarrow\define\sup_{t\in[t_i, t_{i+1}]} S_t, \quad i\in\{1,\dots,n\}.
    \end{equation*}
    \cite{Parkinson_estimator} introduces an extreme value estimator, henceforth called the \emph{Parkinson estimator}:
    \begin{equation*}
        \sqrt{\frac{1}{4 n \ln 2} \sum_{i=1}^n \left(\ln\frac{S_i^\uparrow}{S_i^\downarrow}\right)^2}.
    \end{equation*}

    \paragraph{Rogers--Satchell}   The Parkinson estimator assumes that the underlying process follows a geometric Brownian motion with zero drift, which is not always the case in real markets. Particularly, during periods when the asset trends strongly, these estimators then overestimate volatility. From this observation, Rogers and Satchell proposed in 1991 a new estimator that allows for non-zero drift \cite{RogersSatchell}.
    The Rogers--Satchell volatility estimator includes opening and closing prices:
    \begin{equation*}
        \sqrt{\frac 1n\sum_{i=1}^n \ln\frac{S_i^\uparrow}{S_{t_{i+1}}} \ln\frac{S_i^\uparrow}{S_{t_{i}}} + \ln \frac{S_i^\downarrow}{S_{t_{i+1}}}\ln\frac{S_i^\downarrow}{S_{t_{i}}} }
    \end{equation*}

    \section{Data and Code}\label{sect:data and code}\label{sect:datasets}\label{sect:code}

    To test the robustness and universality of the stylized facts we consider, we have included a diverse range of assets, asset classes, and exchange regions in our analysis. 
    We include both traditional financial market data and cryptocurrency data to compare their respective characteristics. 
    Below are the specific datasets used:

    \begin{itemize}
        \item \textbf{Minute data:} High-frequency (minute) trading data from futures markets.
        This data allows the investigation of behaviors in short time frames and includes 
        Soybeans, Crude Oil, Nikkei, Japanese Yen, FTSE, British Pound, Euro FX, S\&P-mini from the beginning of 2012 up to the end of 2022.  
        \item \textbf{Daily data:} Approximately 10 years of daily futures market data consisting of Canadian 10-year bond, GB long gilt, Switzerland stock exchange, Austrian stock exchange, Korean stock exchange, Euro, USD, and Chinese Yuan to Swiss francs, along with gold and live cattle futures from February 2022 until the end of 2023. This dataset covers traditional market instruments with sufficient granularity to study medium-term phenomena.
        \item \textbf{Monthly data:} Long-term financial data dating back to 1871 for various asset classes and regions. This historical overview enables the examination of stylized facts over extended horizons. This dataset consists of copper, Texas oil, wheat, GB, US, and JP 10-year bonds, as well as exchange rates of CHF, GBP, and Japanese Yen to USD. 
        \item \textbf{Cryptocurrency data:} minute data for Bitcoin and Ethereum from the beginning of 2017 up to the end of 2022. As cryptocurrencies operate in a continuous 24-hour market without specific open/high/low conventions, additional pre-processing steps are conducted to proxy these measures. 
    \end{itemize}

    Data for cryptocurrencies and classical daily data is sourced through Google Finance\footnote{\url{https://www.google.com/finance}}. Long-term historical data in monthly scales, as well as intraday data for futures markets, are retrieved from private databases such as Global Financial Data\footnote{\url{https://globalfinancialdata.com}} and Tick Data\footnote{\url{https://www.tickdata.com}}. While the first is freely accessible to the public, the latter is proprietary and restricted by private contracts.

    Our experiments are implemented in a publicly accessible \texttt{Python} \cite{python} pipeline hosted on \texttt{GitHub}\footnote{\url{https://github.com/Saraaqzs/stylized\_dev}} to facilitate reproducibility. This pipeline leverages several open-source libraries \cite{2020SciPy-NMeth,harris2020array,mckinney2010data,seabold2010statsmodels,Hunter:2007}.

    \section{Statistical tests and results}
    \label{sect:Results}
     
    In this section, we explain the statistical tests we used for each stylized fact, followed by the experimental result of the tests on the different datasets described in Section~\ref{sect:datasets}. 
    
    \subsection{Absence of autocorrelations for log-returns}\label{sect:absence-of-autocorrelations}
    
    \paragraph{Description} Let $\delta$ be the time step of the returns data, for example one minute for minute returns, or one day for daily returns. 
    This stylized fact asserts that for two times $t,\tilde t\in I$, the correlation coefficients between $r_{t,t+\delta}$ and $r_{\tilde t, \tilde t+\delta}$ are close to $0$ for $\delta<\tilde t - t$ and when $\tilde t - t$ is larger than roughly 20 minutes. 

    To test for autocorrelation between logarithmic returns, we investigate their autocorrelation function, henceforth denoted by ACF.
    \begin{equation*}
        \operatorname{ACF}(l)=\operatorname{ACF}(l,\delta,t)\define \operatorname{Corr}(r_{t,t+\delta}, r_{t+l,t+l+\delta}).
    \end{equation*}
    Here, $l$ is the time lag (typically a multiple of $\delta$) and $t$ is the absolute time at which we define the covariance.

    \paragraph{Statistical test}
    Since in practice we only have a single realization of the time series, we estimate the ACF by assuming the stationarity of the process. Under this assumption, we estimate $\operatorname{ACF}(l)$ using an average over all time windows of length $l$. Specifically, for each lag $l$, we calculate first the auto-covariance estimator
    \begin{equation*}
    \widehat{\operatorname{AutoCov}}(l) = \frac{1}{|T|-1} \sum_{t \in T} \left( r_{t,t+\delta} - \bar{r} \right)\left( r_{t+l,t+l+\delta} - \bar{r} \right),
    \end{equation*}
    where $T$ is the set of times over which the summation is performed (assumed to be non-empty), $|T|$ is the cardinality of $T$, and $\bar{r}$ is the (empirical) mean of the log returns over the time series, given by:
    \begin{equation*}
    \bar{r} = \frac{1}{|T|} \sum_{t \in T} r_{t,t+\delta}. 
    \end{equation*}
    We then normalize the auto-covariance by the estimated variance ($\widehat{\operatorname{AutoCov}}(0)$) of the returns (using the basic estimator from Section~\ref{sect:volatility}) to obtain the auto-correlation function. That is, the estimated ACF is given by:
    \[ \widehat{\operatorname{ACF}}(l) = \frac{\widehat{\operatorname{AutoCov}}(l)}{\widehat{\operatorname{AutoCov}}(0)} . \]
    To assess the statistical significance of the estimated ACF values, we compute confidence intervals with Bartlett's formula (\cite[Section 7.2]{tsAnalysis_BrockwellDavis},\cite{seabold2010statsmodels}). This enables us to test whether the observed values differ significantly from 0, indicating the presence of autocorrelations. 
     
\begin{figure}
    \centering
    \includegraphics[scale=0.33]{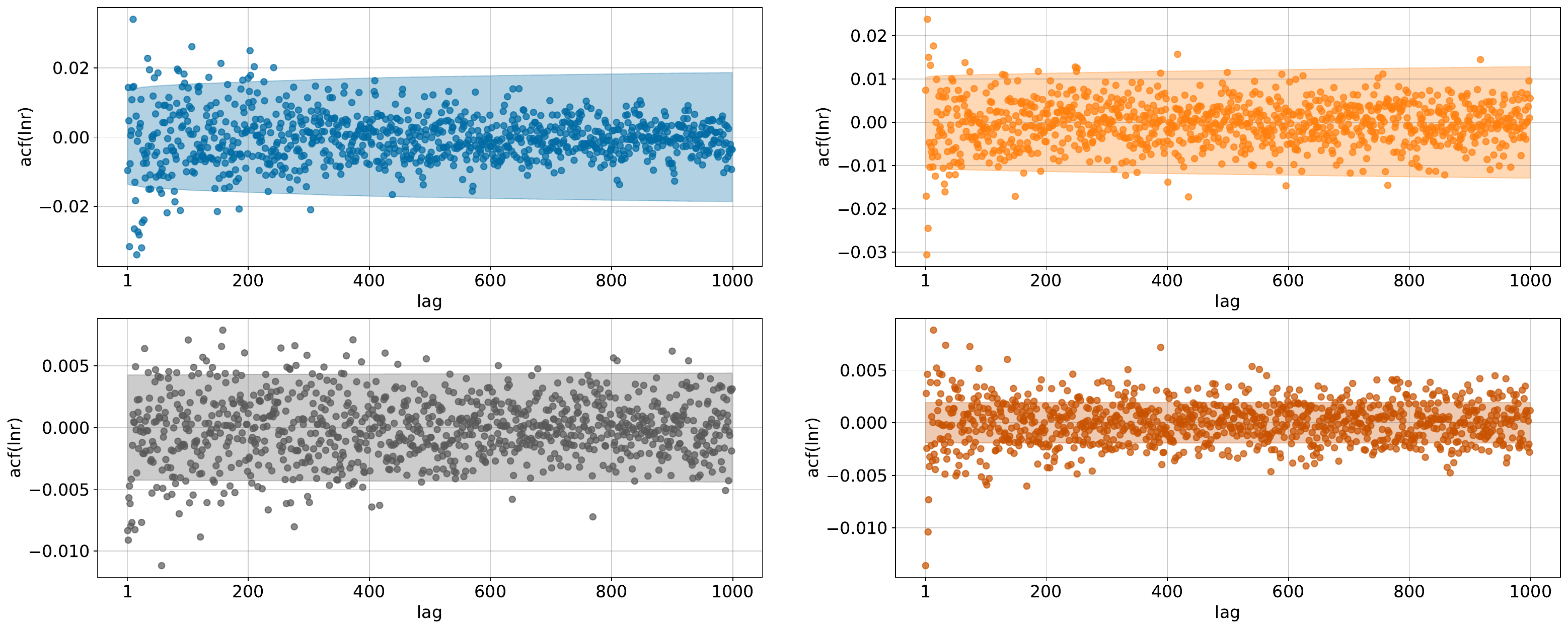}
    \caption{Averaged autocorrelation of the log-returns versus lag (Section~\ref{sect:absence-of-autocorrelations}) in different time scales of monthly (blue), daily (orange), 1-minute (gray), and 1-minute cryptocurrencies (red) over the assets explained in section ~\ref{sect:datasets}. The 95\% confidence interval of cryptocurrencies looks tighter than traditional markets.}
    \label{fig:absence_acf}
    \end{figure}

    \paragraph{Result} 
    We compute the empirical ACFs and their 95\% confidence intervals for individual assets, including high-frequency cryptocurrency data to establish a comparison with traditional markets (see Figure~\ref{fig:absence_acf}). For brevity, we report only the mean ACF -- averaged over the series present in each data group which is mentioned in section ~\ref{sect:datasets} -- graphically.

\begin{figure}
    \centering
    \includegraphics[scale=0.33]{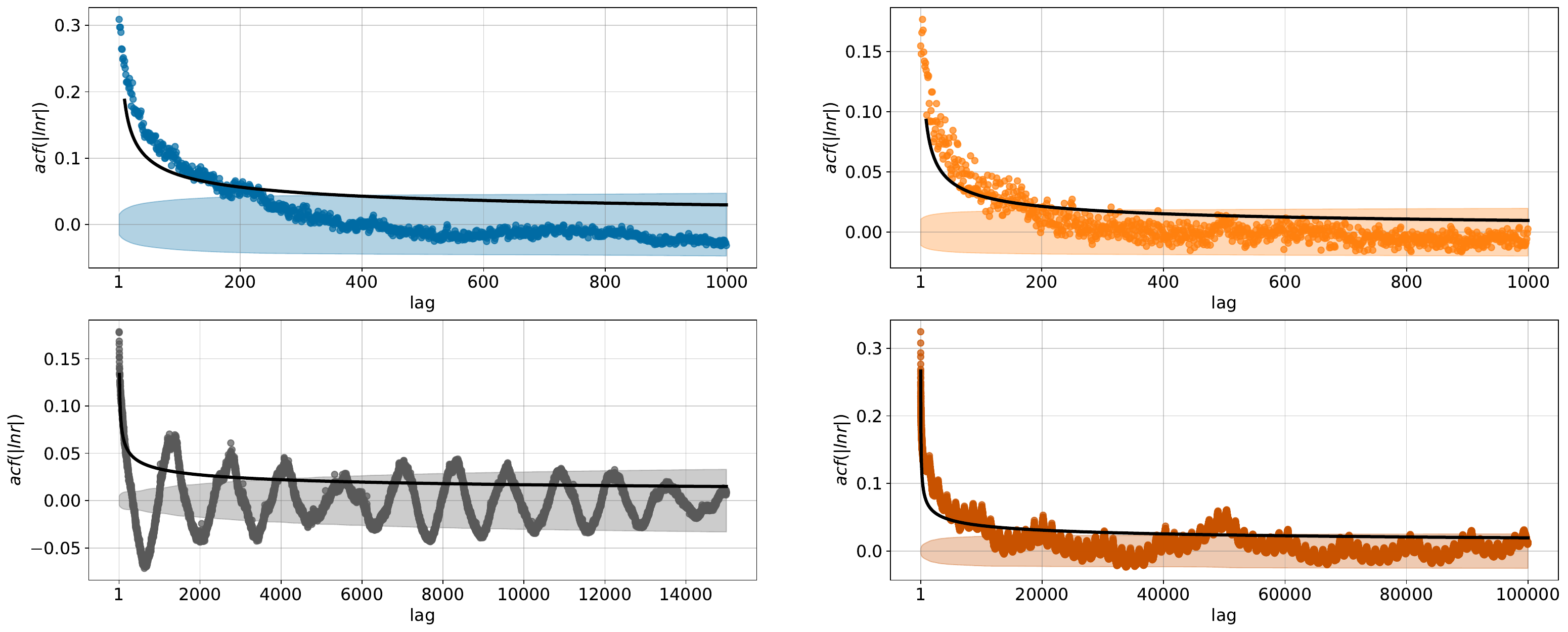}
        \includegraphics[scale=0.33]{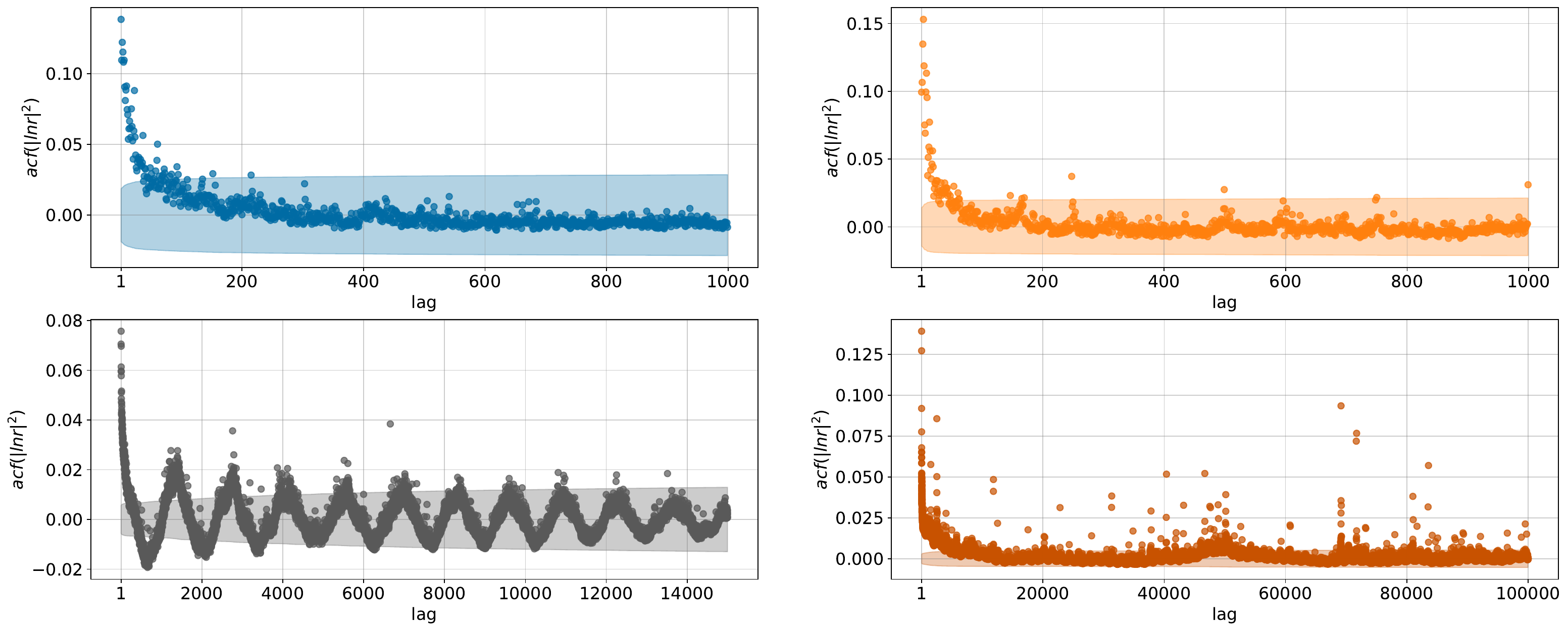}

    \caption{Averaged autocorrelation functions (ACF) for absolute log-returns (upper four panels) and squared absolute log-returns (lower four panels), corresponding to Section~\ref{sect:slow-decay-of-autocorrelations}. Results are shown for different time scales and asset types: monthly (blue), daily (orange), 1-minute traditional markets (gray), and 1-minute cryptocurrencies (red). The solid black curves in the upper panels represent power-law fits ($l^{-\beta}$) to the ACF of absolute log-returns, as detailed in Section~\ref{sect:slow-decay-of-autocorrelations}. The fitted exponents $\beta$ are reported quantitatively in Table~\ref{tab:beta}.}
    \label{fig:slow_acf}
    \end{figure}

    As shown in Figure~\ref{fig:absence_acf}, the log-return autocorrelation functions quickly become consistent with zero within the 95\% confidence interval for all groups of datasets, except for the cryptocurrencies. This is more prominent when contrasted with the pattern of decay detailed in section~\ref{sect:slow-decay-of-autocorrelations}, figure~\ref{fig:slow_acf}. Our finding is not surprising as for the traditional financial market it is well established and given that cryptocurrencies operate under different pricing mechanisms and market structures compared to classical financial markets, we hypothesized that their autocorrelation patterns might exhibit distinctive characteristics.

    A slow decay of the autocorrelation function (ACF) of a time series indicates a long-range dependence in the data \cite{Cont_longRange}. Initial studies from the 1930s and 1950s \cite{working1934, kendall1953} found that financial time series exhibit ACF values close to zero and reported the serial correlation to be negligible for any effective practical usage, within the confidence interval.
    For high-frequency data, however, significant non-zero autocorrelation has been reported \cite{GoodhartFig,andersen_boll1997,Gency2001} and documented by \cite{Cont2001} (see also \cite{Cont_Beyond, Stylized_modern}). This autocorrelation is typically negative in the first few lags, and often attributed to the bid-ask bounce effect often observed in market data. Further dependencies in tick data have also been observed by \cite{zhou1996}.

\begin{table}
\centering
\begin{tabular}{ |p{3.5cm}||p{2.5cm}|p{2.5cm}|p{2.5cm}| }	\hline
 	Asset Class & $\beta \pm \Delta\beta$\\ \hline
    \multicolumn{2}{|l|}{Traditional Markets}  \\	\hline
	Monthly  & $0.38 \pm 0.0070$ \\  
	Daily & $ 0.45 \pm 0.073 $ \\	
    Intraday & $ 0.33 \pm  0.0071 $ \\	 
 \hline \multicolumn{2}{|l|}{Cryptocurrencies}  \\	\hline
	Intraday & $ 0.28 \pm 0.0035 $ \\ \hline
\end{tabular}
\caption{Estimated power-law decay exponents $\beta$ for autocorrelation of absolute log-returns across different time scales and market types, along with their one standard deviation uncertainties, $\Delta\beta$, as described in section \ref{sect:absence-of-autocorrelations}. }
\label{tab:beta}
\end{table}

    \subsection{Slow decay of autocorrelations for absolute log-returns}\label{sect:slow-decay-of-autocorrelations}
    \paragraph{Description} 
    This stylized fact asserts that for $t,\tilde t \in I$, the correlation coefficients between $\abs{r_{t,t+\delta}}$ and $\abs{r_{\tilde t, \tilde t+\delta}}$ decay slowly for growing $\tilde t-t$ and a fixed time step $\delta>0$, in contrast with the correlation coefficients for the log-returns themselves, which are hypothesized to decay quickly (see Section~\ref{sect:absence-of-autocorrelations}). It is suggested in \cite{Cont2001} that the correlation coefficient, denoted by $\Corr$, between the two behaves roughly as
    \begin{equation}\label{eq:polynomial}
        \Corr(\abs{r_{t,t+\delta}}^{\alpha}, \abs{r_{\tilde t, \tilde t+\delta}}^{\alpha})\approx (\tilde t - t)^{-\beta}, \qquad\text{as }\tilde t - t\to\infty,
    \end{equation}
    for some $\beta\in[0.2,0.4]$, where we will use $\alpha=1$ and $\alpha=2$. It is furthermore asserted that the exponent $\beta$ is roughly independent of $\delta$. 

    \paragraph{Statistical test}
    We estimate the autocorrelation function and obtain standard errors for our estimate as in Section~\ref{sect:absence-of-autocorrelations}, with the difference that we now investigate the correlation function of the absolute value of the logarithmic returns, instead of the logarithmic returns themselves.
    To fit the exponent $\beta$, we consider the family of functions
    \begin{equation*}
    l\mapsto l^{-\beta},
    \end{equation*}
    and look for the $\beta$ which minimizes the least-square objective
    \begin{equation*}
    \sum_{l \in \mathcal{L}} \left( \widehat{\operatorname{ACF}}(l) - l^{-\beta} \right)^2,
    \end{equation*}
    where, as before, $\widehat{\operatorname{ACF}}(l)$ denotes the estimated autocorrelation function of absolute log-returns at lag $l$, and $\mathcal{L}$ is the set of lags considered in the estimation. 
    We numerically search for the optimal $\beta$ using the Levenberg-Marquardt algorithm provided by Scipy \cite{2020SciPy-NMeth}.

   The one-standard-deviation errors $\Delta\beta$ reported in Table~\ref{tab:beta} are estimated from the covariance matrix derived during the Levenberg-Marquardt optimization \cite{JJ_scipy_opt}. This estimation assumes that the residuals (the differences between the empirical ACF values $\widehat{\operatorname{ACF}}(l)$ and the fitted power-law $l^{-\beta}$) are approximately normally distributed with a constant variance $\sigma^2$. Under this assumption, the variance for the estimated parameter $\beta$ is approximated by
    \begin{equation*}
    \widehat{\operatorname{Var}}(\beta) \approx \hat{\sigma}^2 (J^TJ)^{-1}_{[\beta, \beta]},
    \end{equation*}
    where $\hat{\sigma}^2$ is the estimated variance of the residuals, obtained via maximum likelihood estimation, $J$ is the Jacobian matrix of the fitting function with respect to $\beta$ evaluated at the optimal parameter value, and $(J^TJ)^{-1}_{[\beta, \beta]}$ denotes the diagonal element corresponding to $\beta$ in the matrix $(J^TJ)^{-1}$. The standard error is then $\Delta\beta = \sqrt{\widehat{\operatorname{Var}}(\beta)}$.
    
\begin{figure}
    \centering
    \includegraphics[scale=0.33]{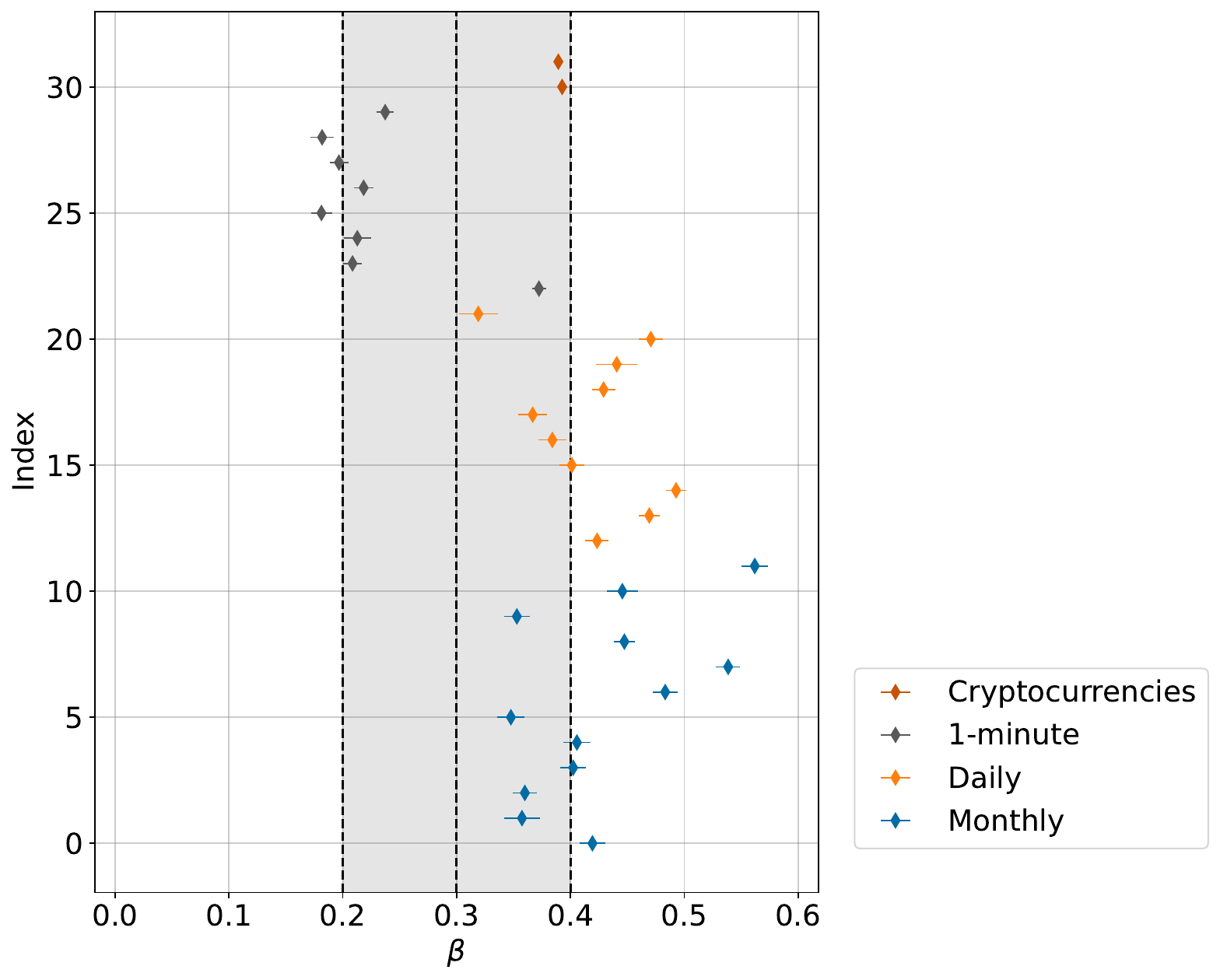}
    \caption{Power-law decay exponents $\beta$ for the autocorrelation of absolute log-returns across different markets, as described in \eqref{eq:polynomial}. The range $[0.2, 0.4]$ into which the exponents are conjectured to fall are highlighted with the shaded region. Each point represents a market with corresponding error bars indicating estimation uncertainty. The vertical axis represents the index number of each market in the database, while the horizontal axis shows the $\beta$ values.}
    \label{fig:betas}
    \end{figure}

    \paragraph{Result} In the same manner as previous section (section ~\ref{sect:absence-of-autocorrelations}), we report the averaged autocorrelations for the cases $\alpha = 1$, and $\alpha=2$ in \eqref{eq:polynomial} averaged over the assets described in Section~\ref{sect:datasets}. To assess the goodness of fit, we plot the fitted curve on top of the ACFs, Figure~\ref{fig:slow_acf}. The quantitative result for the exponents $\beta$, both for the averages acfs and per-assets ones, can be found in Table~\ref{tab:beta} and Figure~\ref{fig:betas}, respectively. They suggest that the autocorrelation of absolute log-returns in traditional markets and cryptocurrency markets decay approximately like a power-law function. As indicated, the average values of $\beta$ lie rather consistently within the interval $[0.2, 0.4]$, excluding some exceptions among the individual assets that lie in $[0.4, 0.5]$. Since the sample size for daily and monthly data in terms of number of assets is not large enough, we hesitate to draw a final conclusion regarding individual assets, leaving the matter open for further investigations. We rather assert that as far as our measurement is concerned the average market behavior is aligning well with the assertions proposed in \cite{Cont2001}, and thus providing evidence in favor of the \enquote{slow decay of autocorrelations for absolute log-returns} stylized fact.

    \subsection{Intermittency}\label{sect:intermittency}
    
    \paragraph{Description} Intermittency is a term coming from the mathematical discipline of dynamical systems, describing the behavior of a system that alternates between different states with distinct observable characteristics. For financial time series, this means alternating between periods of high volatility (often associated with crises) and periods of low volatility.
    
    \paragraph{Statistical test} The difficulty to find a precise definition of intermittency in the literature led us to include several, complementary approaches. 
    
    \begin{enumerate}
        \item We investigate the autocorrelation $\operatorname{Cov}(\sigma_t, \sigma_{t+\delta})$ of the volatility of the asset, where $\delta>0$ is the time lag. The fourth stylized fact (volatility clustering, Section \ref{sect:volatility-clustering}) asserts that this autocorrelation is positive, at least for short time lags. We claim that intermittency will lead to a \emph{negative} autocorrelation for intermediate time lags. 
        \item We visually compare the time series with models that are accepted as being intermittent (GARCH$(1,1)$) and others that are accepted as not being intermittent (OU processes) .
        \item We consider the \textit{excursion length} above a certain volatility level. Qualitatively, in an intermittent process that experiences regime changes, the tail events are clustered in a limited period that differentiates from the low volatility period. Due to clustering as outlined above, a high volatility instance of the time series is expected to be followed by another high volatility until the cluster disappears. This introduces shorter excursion length for tail events once happen.
    \end{enumerate}

\begin{figure}
    \centering
    \includegraphics[scale=0.25]{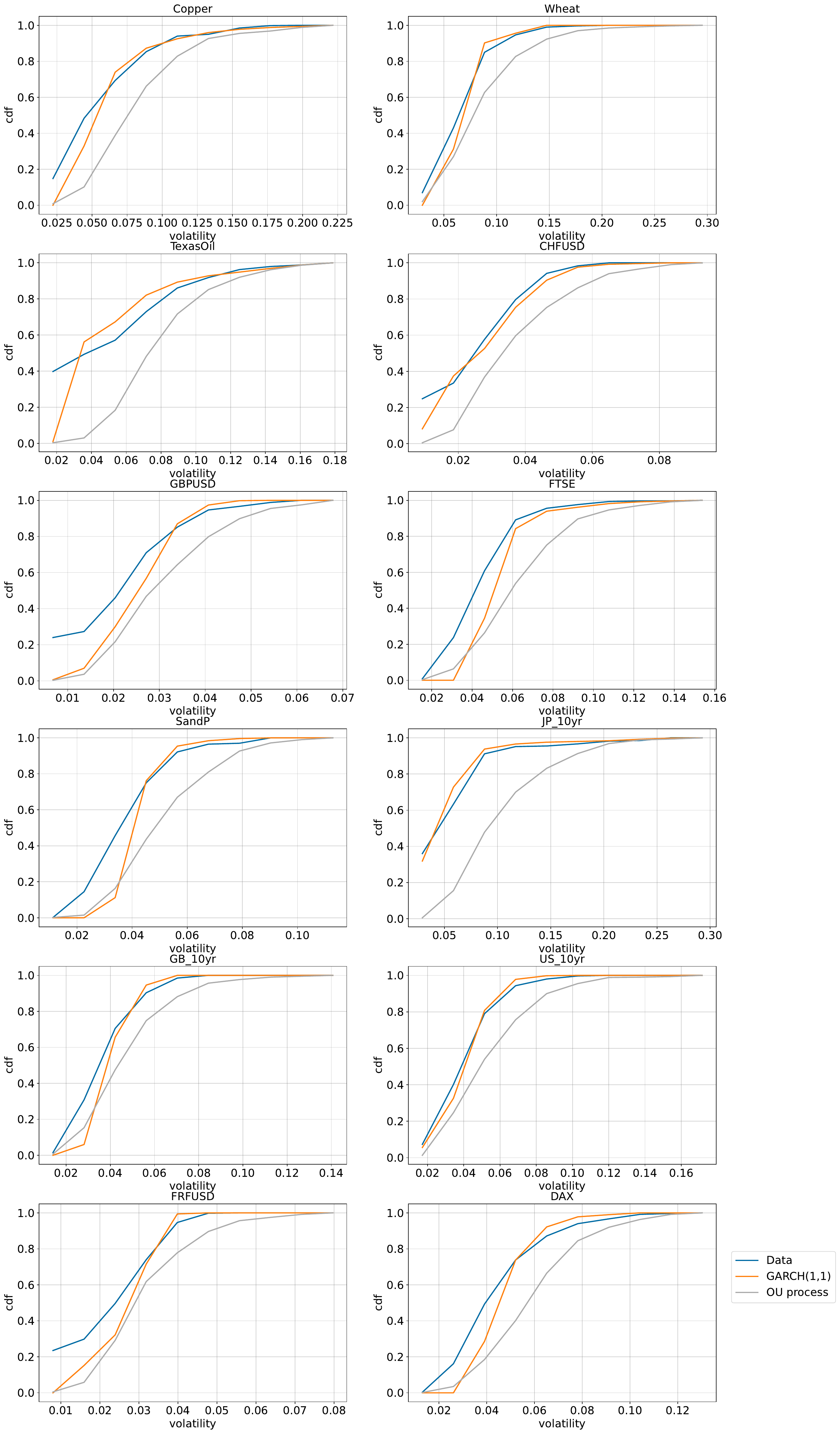}
    \caption{Cumulative distribution function for basic volatility of monthly markets (blue), fitted GARCH(1,1) (orange), and fitted OU (gray). In the large volatility data curve tends to lie between the two curve closer to the GARCH model that captures intermittency well.}
    \label{fig:inter_cdf_m}
    \end{figure}

\begin{figure}
    \centering
    \includegraphics[scale=0.25]{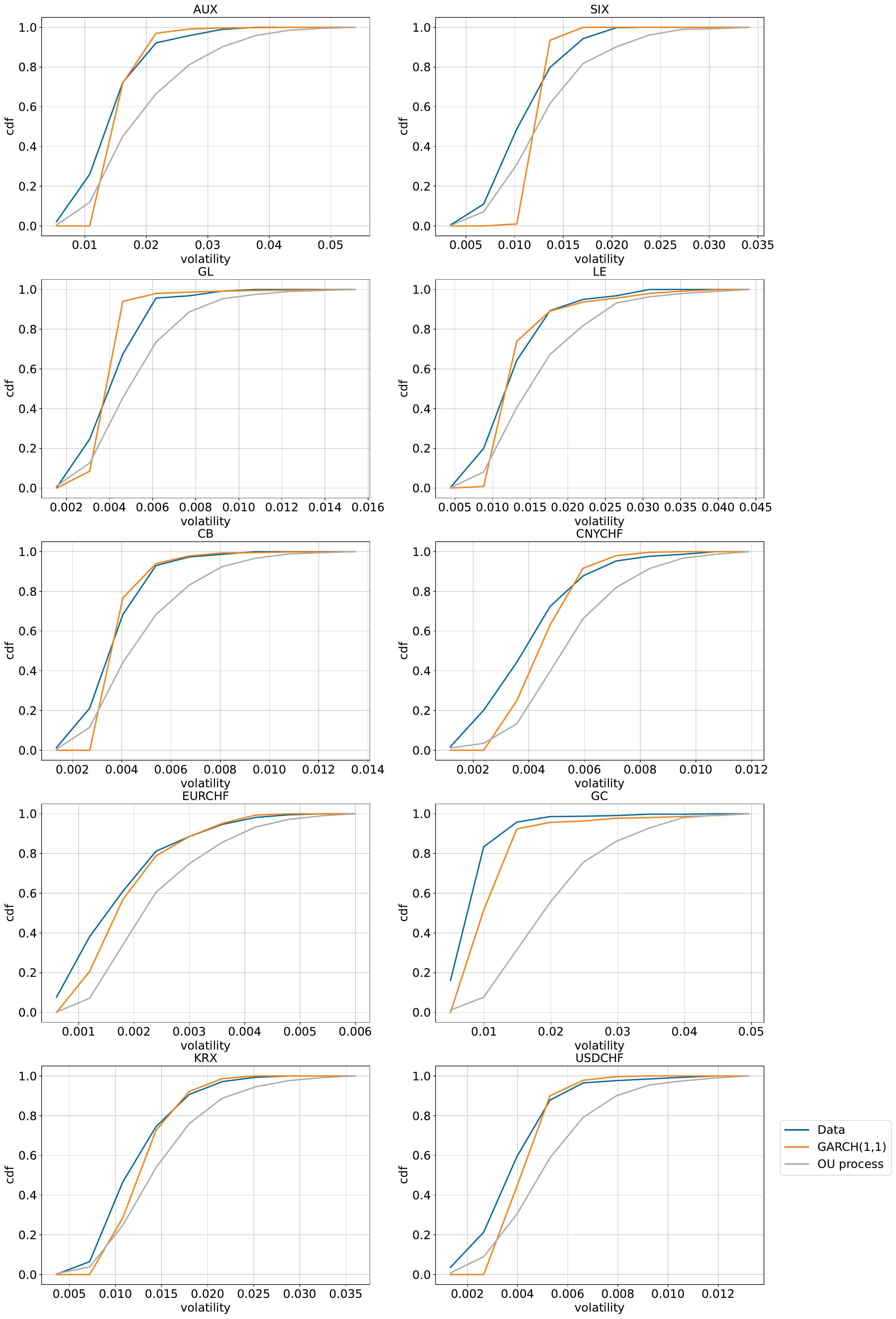}
    \caption{Cumulative distribution function for basic volatility of daily markets (blue), fitted GARCH(1,1) (orange), and fitted OU (gray). In the large volatility data curve tends to lie between the two curve closer to the GARCH model that captures intermittency well.}
    \label{fig:inter_cdf_d}
    \end{figure}

\begin{figure}
    \centering
    \includegraphics[scale=0.3]{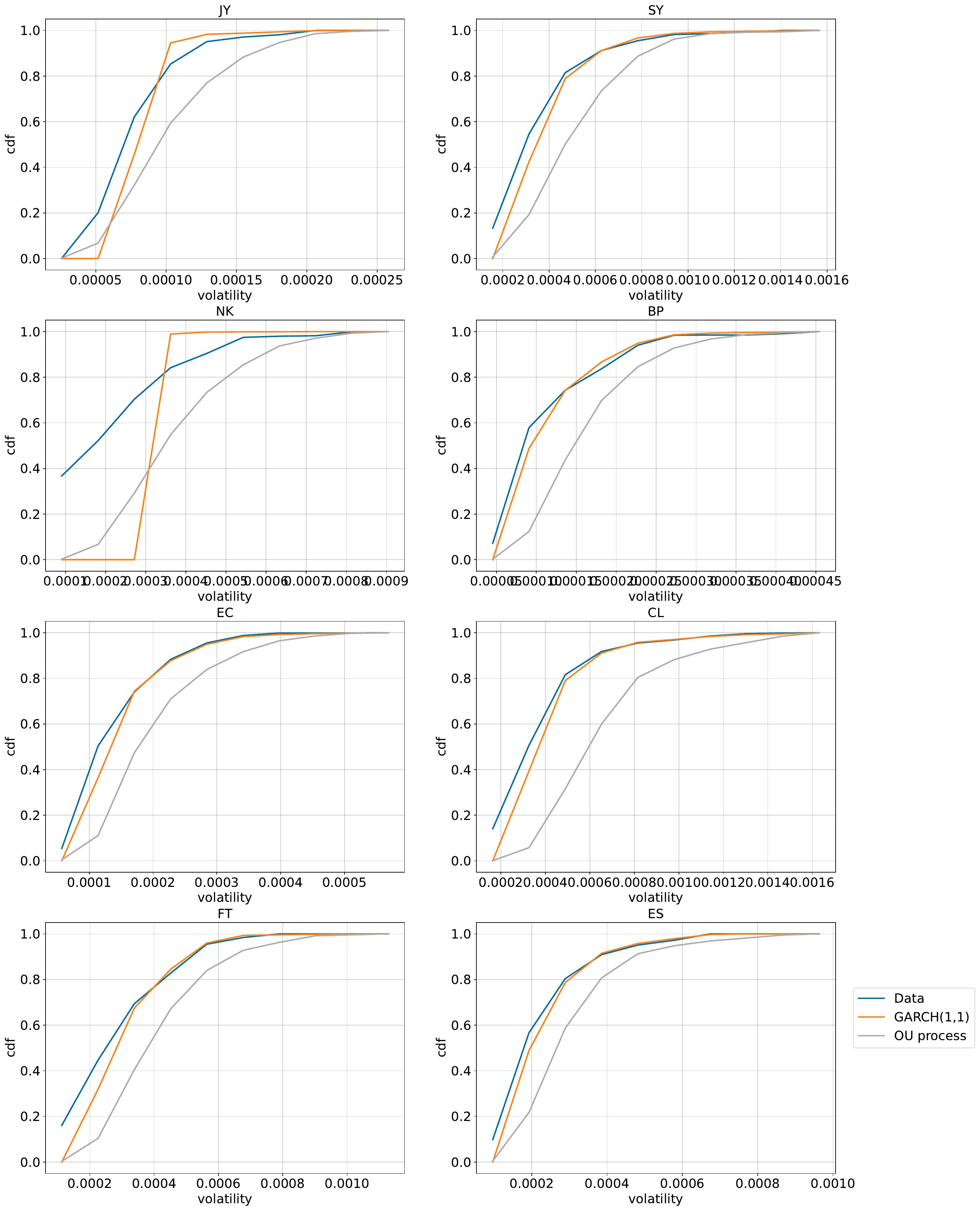}
    \caption{Cumulative distribution function for basic volatility of minute markets (blue), fitted GARCH(1,1) (orange), and fitted OU (gray). In the large volatility data curve tends to lie between the two curve closer to the GARCH model that captures intermittency well.}
    \label{fig:inter_cdf_hf}
    \end{figure}

\begin{figure}
    \centering
    \includegraphics[scale=0.3]{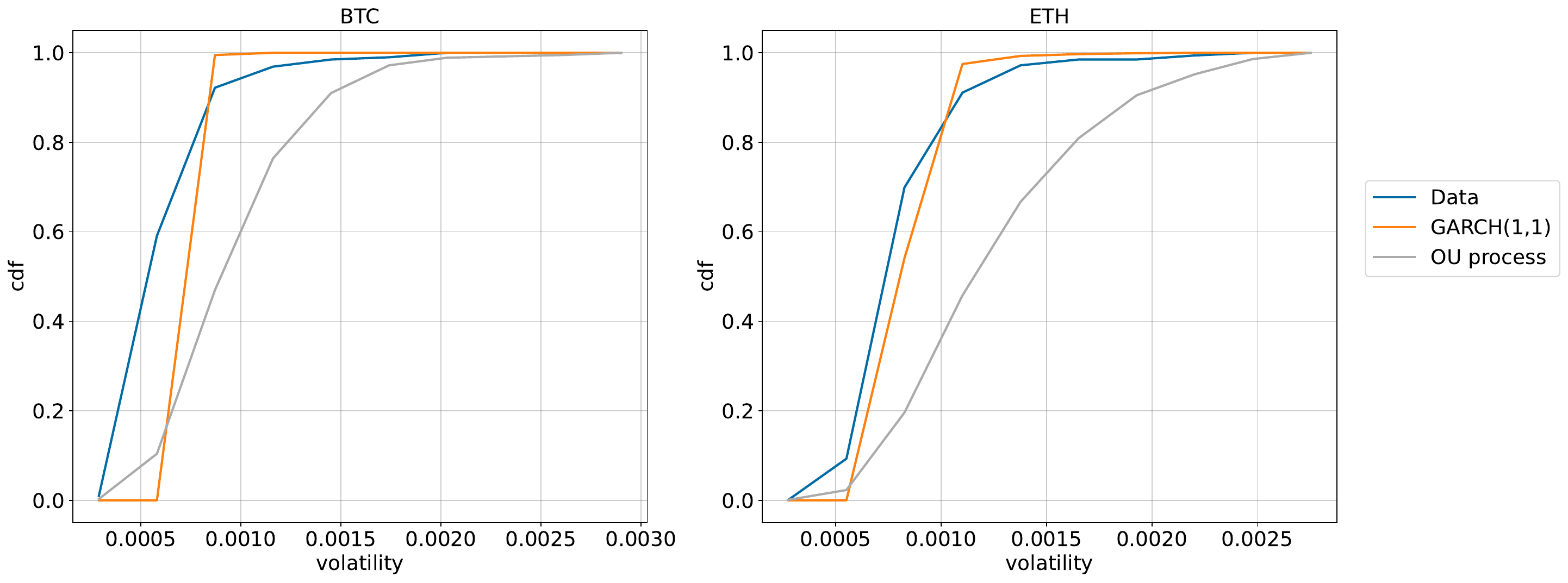}
    \caption{Cumulative distribution function for basic volatility of cryptocurrencies (blue), fitted GARCH(1,1) (orange), and fitted OU (gray). In the large volatility data curve tends to lie between the two curve closer to the GARCH model that captures intermittency well.}
    \label{fig:inter_cdf_cryptos}
    \end{figure}

\begin{figure}
    \centering
    \includegraphics[scale=0.25]{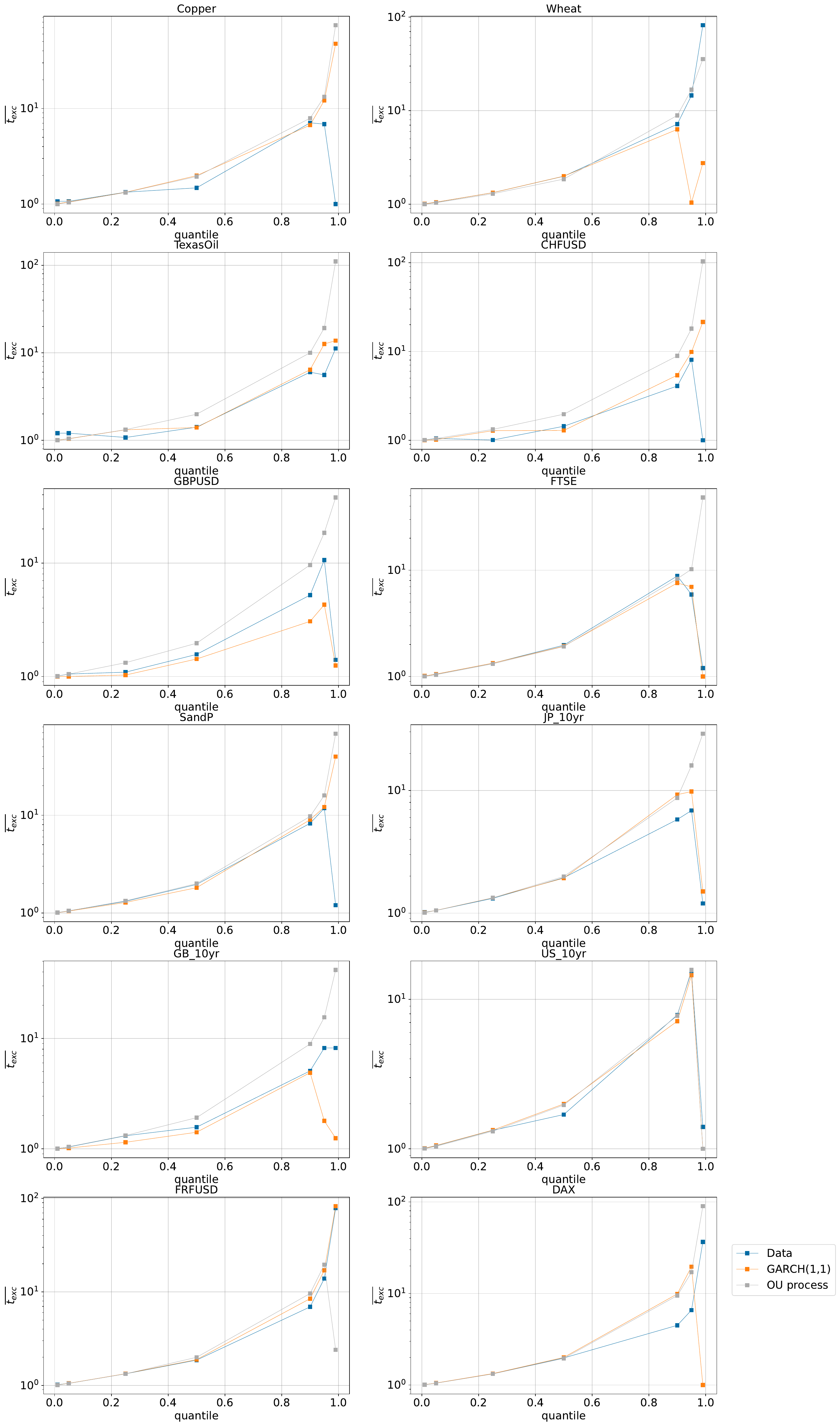}
    \caption{Excursion length of monthly market basic volatility compared to the two aforementioned models. In many of these assets data shows similar behavior as the fiducial intermittent model for the higher quantiles. The plots show the quantiles $1, 5, 25, 50, 90, 95,$ and $99$ percent, respectively.}
    \label{fig:inter_ex_m}
    \end{figure}

\begin{figure}
    \centering
    \includegraphics[scale=0.25]{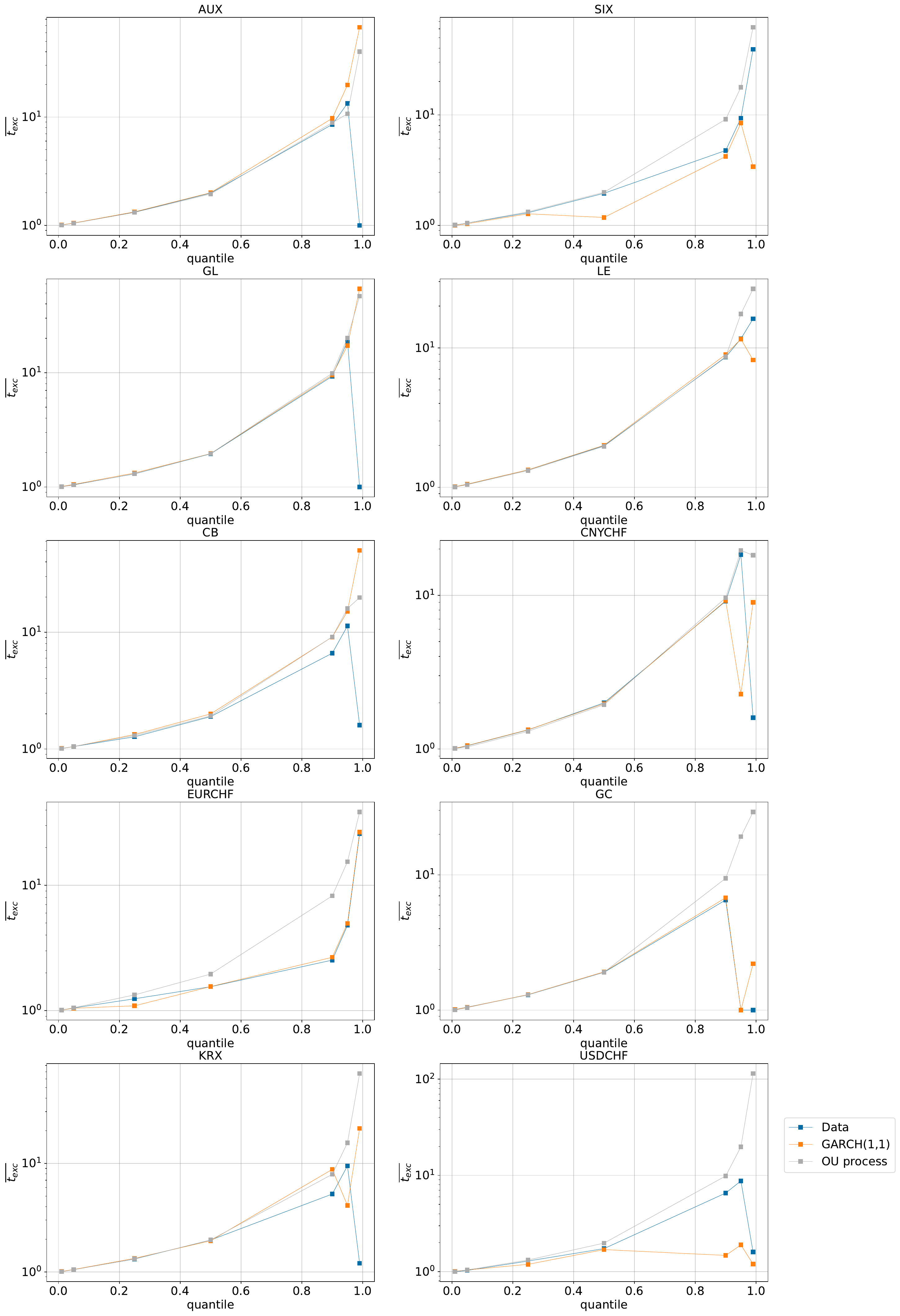}
    \caption{Excursion length of daily market basic volatility compared to the two aforementioned models. In many of these assets data shows similar behavior as the fiducial intermittent model for the higher quantiles. The plots show the quantiles $1, 5, 25, 50, 90, 95,$ and $99$ percent, respectively.}
    \label{fig:inter_ex_d}
    \end{figure}

\begin{figure}
    \centering
    \includegraphics[scale=0.3]{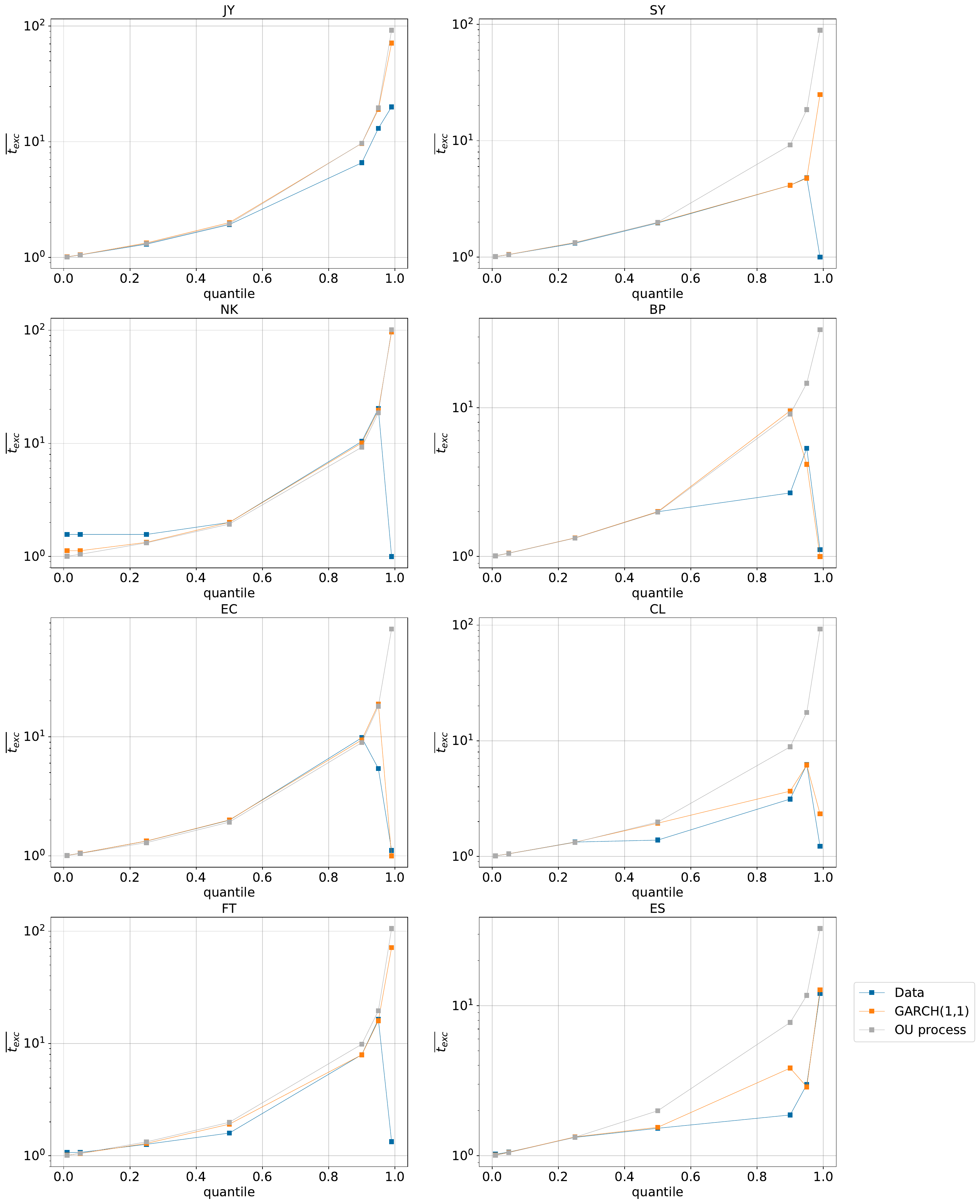}
    \caption{Excursion length of 1-minute market basic volatility compared to the two aforementioned models. In many of these assets data shows similar behavior as the fiducial intermittent model for the higher quantiles. The plots show the quantiles $1, 5, 25, 50, 90, 95,$ and $99$ percent, respectively.}
    \label{fig:inter_ex_hf}
    \end{figure}

\begin{figure}
    \centering
    \includegraphics[scale=0.3]{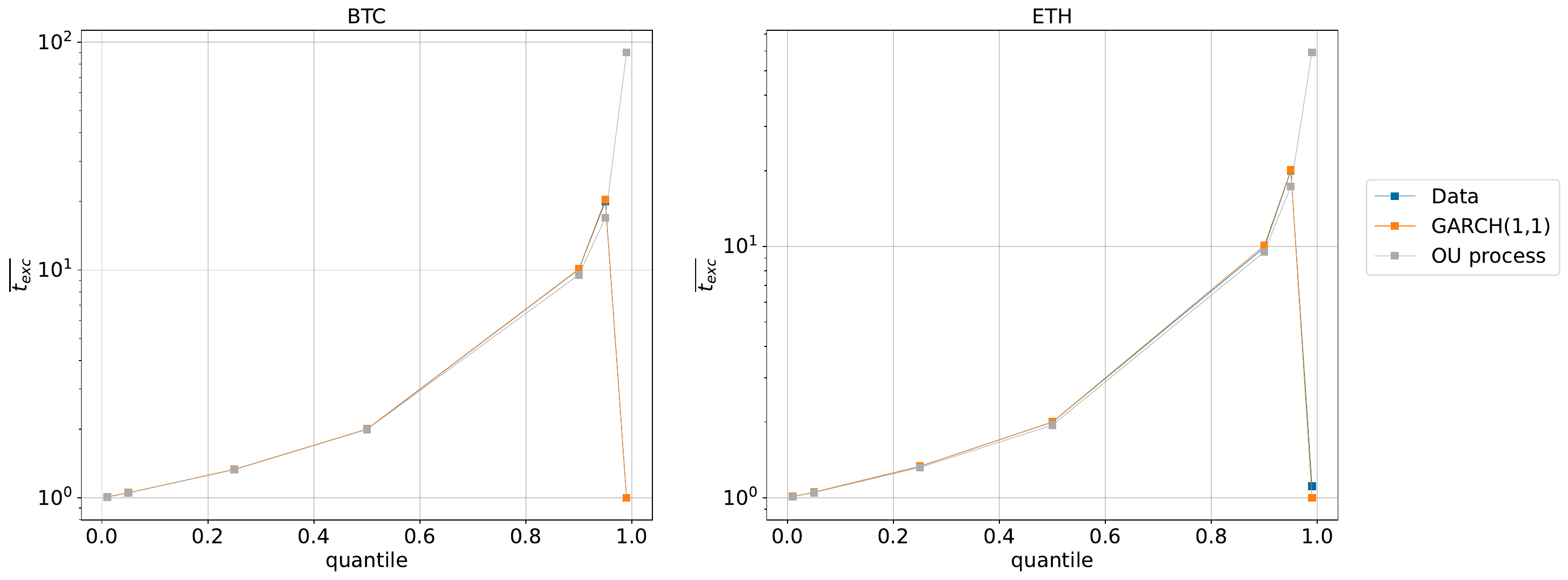}
    \caption{Excursion length of cryptocurrencies basic volatility compared to the two aforementioned models. In many of these assets data shows similar behavior as the fiducial intermittent model for the higher quantiles. The plots show the quantiles $1, 5, 25, 50, 90, 95,$ and $99$ percent, respectively. }
    \label{fig:inter_ex_cryptos}
    \end{figure}

    \paragraph{Result} 
    We now present the results for the intermittency tests outlined previously. One aspect related to intermittency is the autocorrelation of volatility, illustrated in Figures~\ref{fig:volas_acf_mon} to~\ref{fig:volas_acf_cryptos} (discussed further in Section~\ref{sect:volatility-clustering}). However, the large confidence intervals in these figures make it difficult to draw firm conclusions about specific autocorrelation patterns relevant to intermittency from this data alone. For the second test involving comparison to benchmark models, we first needed to fit GARCH models. To do this, we tested the stationarity of the underlying volatility processes using the Augmented Dickey-Fuller (ADF) test \cite{adf}. The test indicated non-stationarity over the full observation period for the long time series, making it necessary to fit the models only to shorter, approximately stationary segments. Figures~\ref{fig:inter_cdf_m} to~\ref{fig:inter_cdf_cryptos} show the cumulative distribution function (CDF) of the empirical basic volatility (Equation~\ref{eq:basic}) for various markets, compared to the CDFs obtained from simulated data using the fitted GARCH(1,1) and Ornstein-Uhlenbeck (OU) processes on these segments. Furthermore, Figures~\ref{fig:inter_ex_m} to~\ref{fig:inter_ex_cryptos} present the results of the excursion time analysis (test 3) for the empirical data and the two benchmark models.
   
   In the initial set of plots, it is evident that for many of the assets we analyzed, the actual volatility data falls between the distributions of the two processes—one known to be intermittent and the other not—with a tendency towards the intermittent process. This is particularly noticeable for higher volatility levels. This observation supports the intermittency stylized fact proposed in \cite{Cont2001}. Additionally, the results for excursion time (Figures~\ref{fig:inter_ex_m} to~\ref{fig:inter_ex_cryptos}) reinforce these findings. We observe that for several assets, once the quantile surpasses a certain threshold in the GARCH model, the excursion time tends to decrease, indicating clustering of high-volatility periods. This behavior is seldom seen in OU models. Furthermore, in many instances, the excursion time for higher quantiles aligns more closely with the behavior of GARCH models than with OU models, suggesting an intermittent time series. As noted, we have only fitted the model to a portion of the time series to maintain stationarity. This sample selection might explain why, in some cases, we do not observe the reversion of the data curve in Figures~\ref{fig:inter_ex_m} to~\ref{fig:inter_ex_cryptos}.

    \subsection{Volatility clustering}\label{sect:volatility-clustering}
    
    \paragraph{Description} This fact asserts that the volatility of an asset is positively auto-correlated over time. Many theoretical models, such as GARCH models or stochastic volatility models like the Heston model \cite{Heston1993}, can capture this behavior of financial markets' time series. Explicitly, the fact states that for small $\delta>0$ and $t\in I$, we have
    \begin{equation*}
        \operatorname{Cov}(\sigma_t, \sigma_{t+\delta})>0.
    \end{equation*}
    This phenomenon has been extensively studied in the literature \cite{Cont_longRange}. 

    \paragraph{Statistical test} To graphically investigate the matter, we use three different metrics of realized volatility described in Section~\ref{sect:volatility}, each of which tailored to be sensitive to certain aspect of the time series volatility, and look at the plots of these metrics all along the available history of the market. Then we compute the empirical lagged autocorrelation which quantitatively suggests the positiveness and the size of this clustering effect per asset.

    \paragraph{Result}
    
\begin{figure}
    \centering
    \includegraphics[scale=0.22]{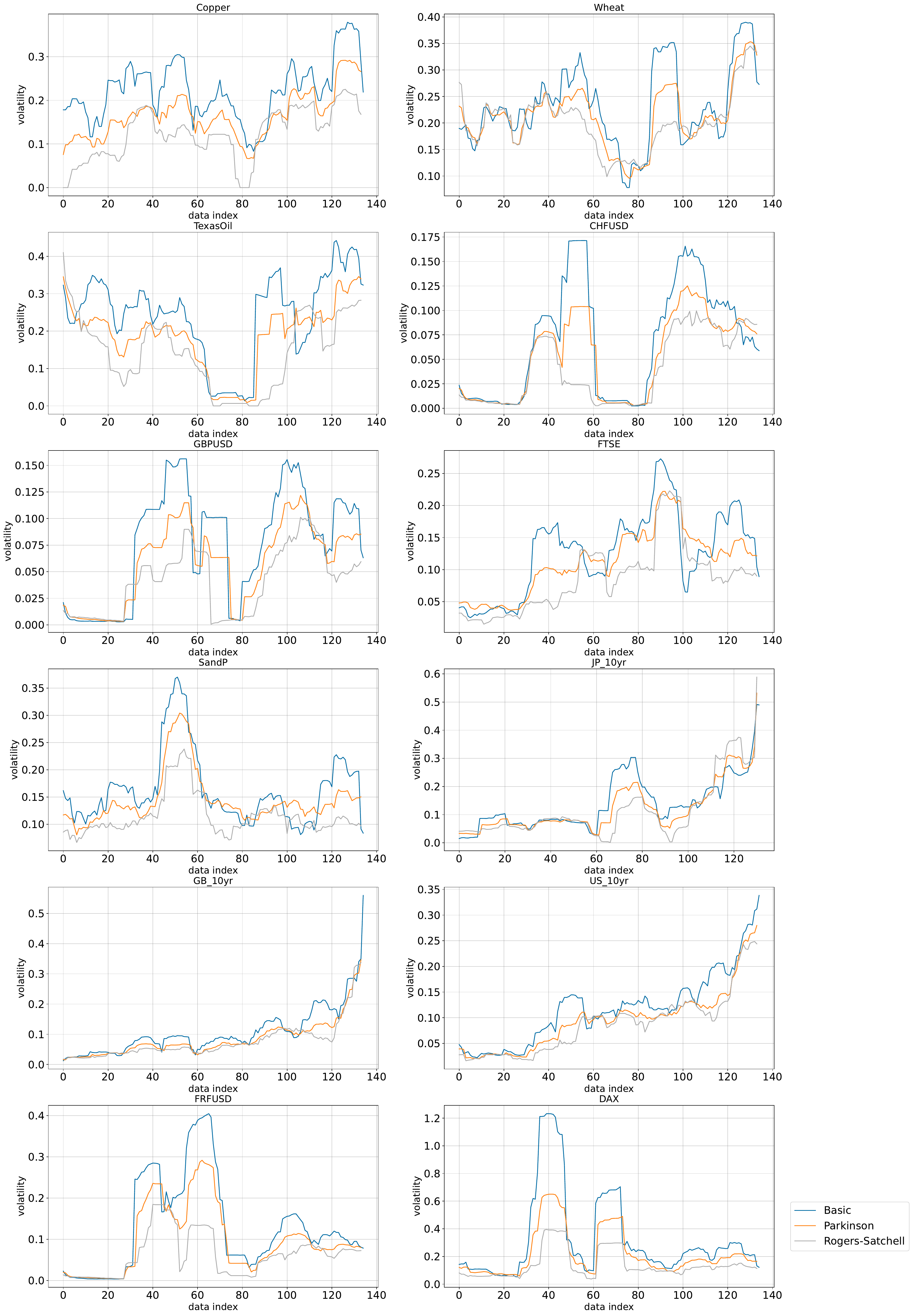}
    \caption{The volatility measures described in Section~\ref{sect:volatility} measured for the monthly dataset of Section~\ref{sect:datasets}. The agreement between different metrics are evident. The horizontal axis shows the  data point index in the time series and the vertical axis is the measure of volatility. The time window over which the historical volatility is computed is constant for all the assets to allow comparison and is 12 months. }
    \label{fig:volas_mon}
    \end{figure}

\begin{figure}
    \centering
    \includegraphics[scale=0.22]{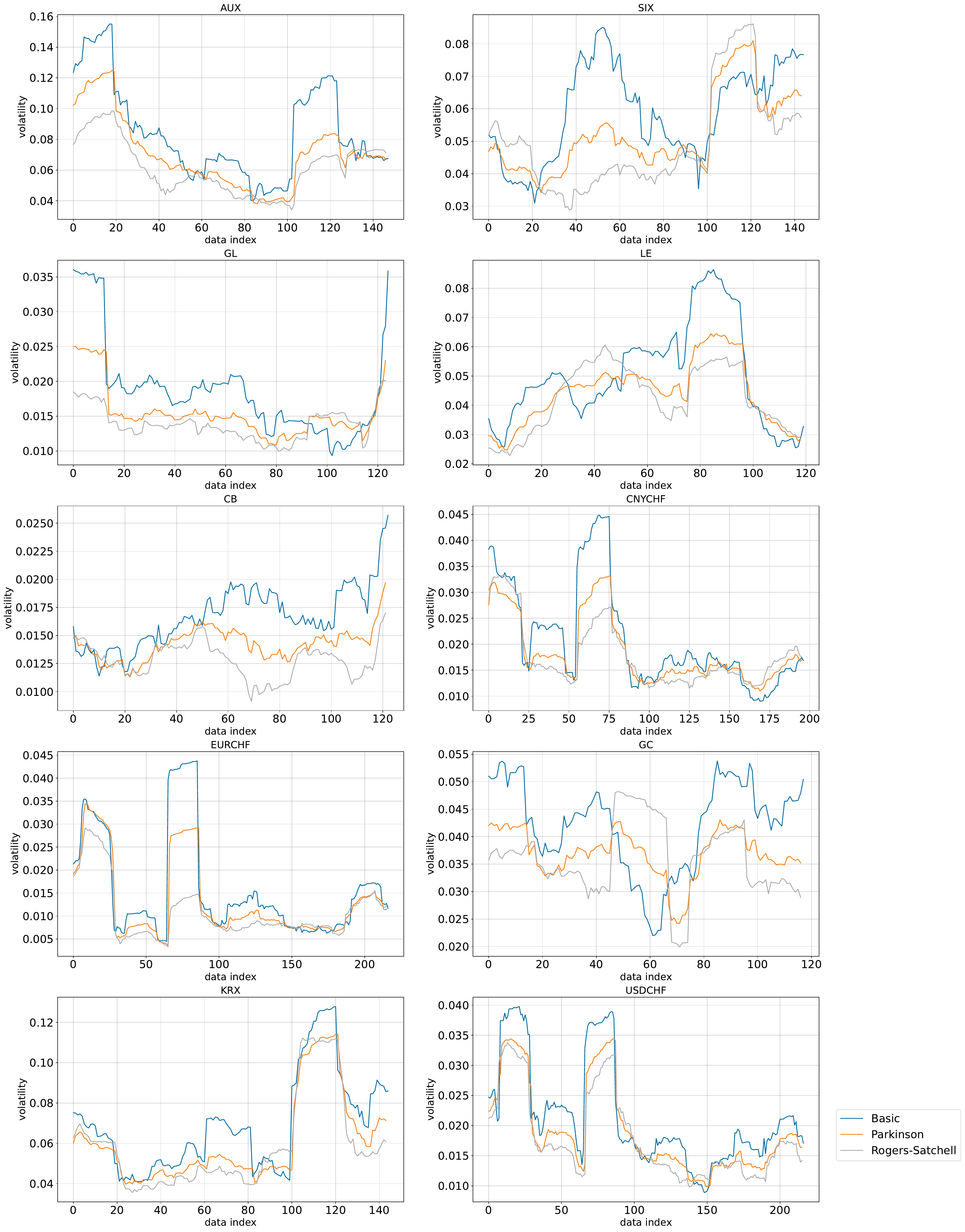}
    \caption{The volatility measures described in Section~\ref{sect:volatility} measured for the daily dataset of Section~\ref{sect:datasets}. The agreement between different metrics are evident. The horizontal axis shows the  data point index in the time series and the vertical axis is the measure of volatility. The time window over which the historical volatility is computed is constant for all the assets to allow comparison and is 21 days.}
    \label{fig:volas_day}
    \end{figure}

\begin{figure}
    \centering
    \includegraphics[scale=0.27]{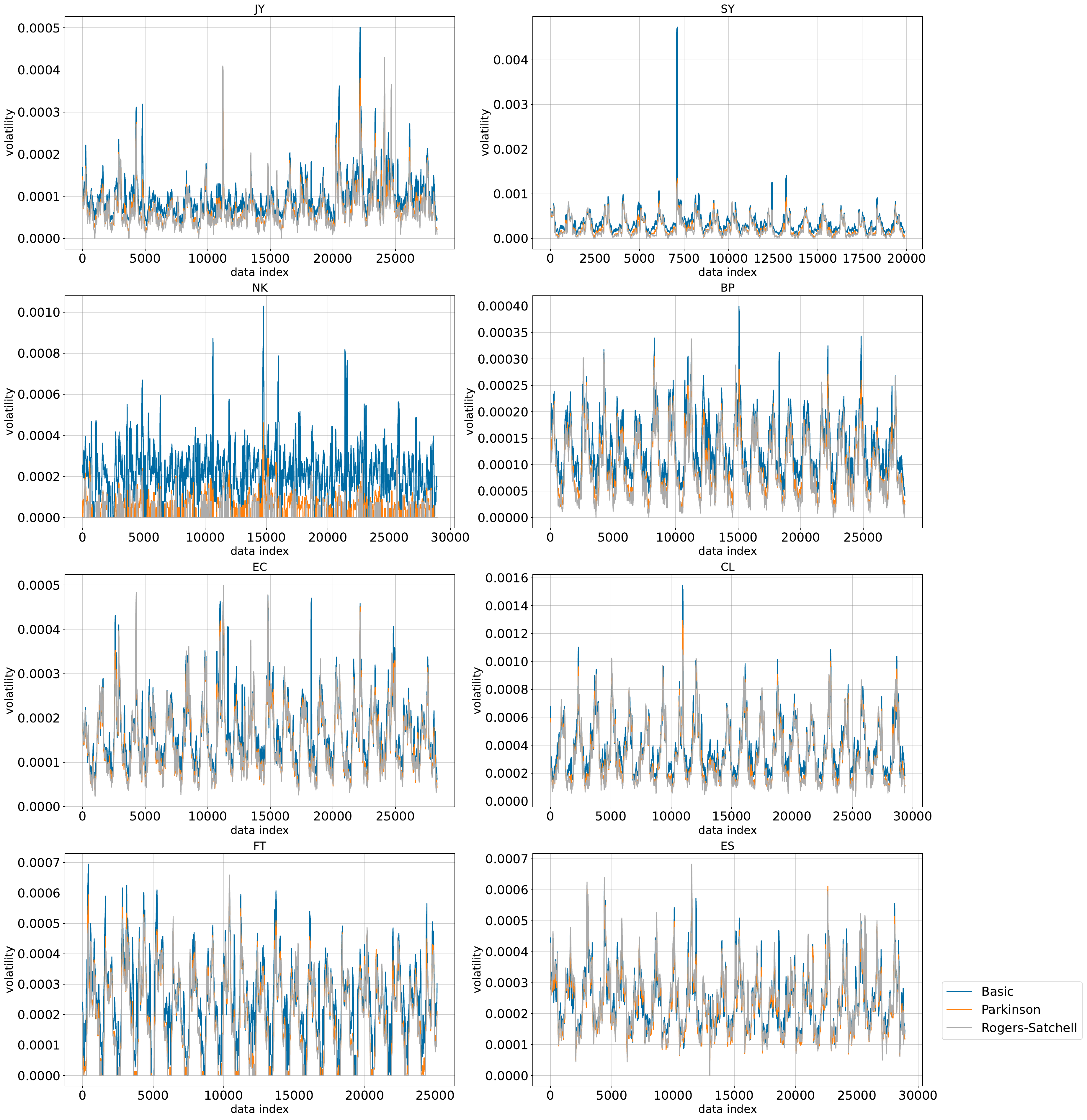}
    \caption{The volatility measures described in Section~\ref{sect:volatility} measured for the high-frequency dataset of Section~\ref{sect:datasets}. The agreement between different metrics are evident. The horizontal axis shows the data point index in the time series and the vertical axis is the measure of volatility. The time window over which the historical volatility is computed is constant for all the assets to allow comparison and is 1 day. One could also see the relative periodicity in the volatility of the high frequency prices induced by scheduled technical underlying events. Also standard deviation captures the daily fluctuations in closing prices, which can be more volatile than the range between high and low prices used in Parkinson measure or the intra-day movements considered in Rogers-Satchell measure. This explains the higher measured basic volatility for some assets such as Nikkei. }
    \label{fig:volas_hf}
    \end{figure}

\begin{figure}
    \centering
    \includegraphics[scale=0.27]{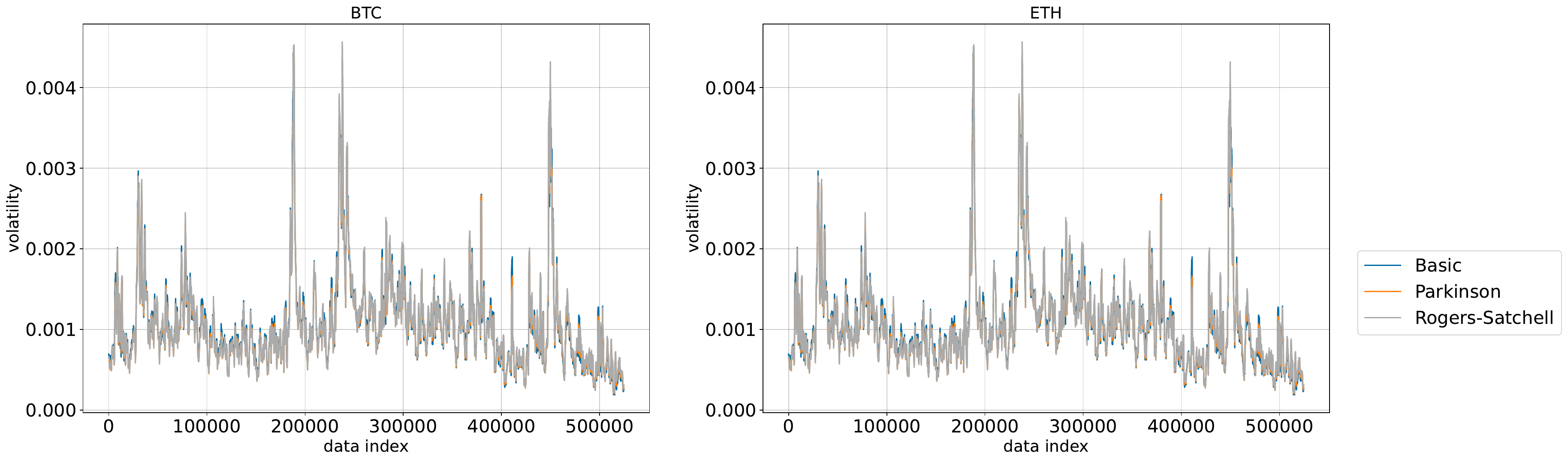}
    \caption{The volatility measures described in Section~\ref{sect:volatility} measured for the cryptocurrency dataset of Section~\ref{sect:datasets}. The agreement between different metrics are evident. The horizontal axis shows the data point index in the time series and the vertical axis is the measure of volatility. The time window over which the historical volatility is computed is constant for all the assets to allow comparison and is 1 day.}
    \label{fig:volas_cryptos}
    \end{figure}

\begin{figure}
    \centering
    \includegraphics[scale=0.22]{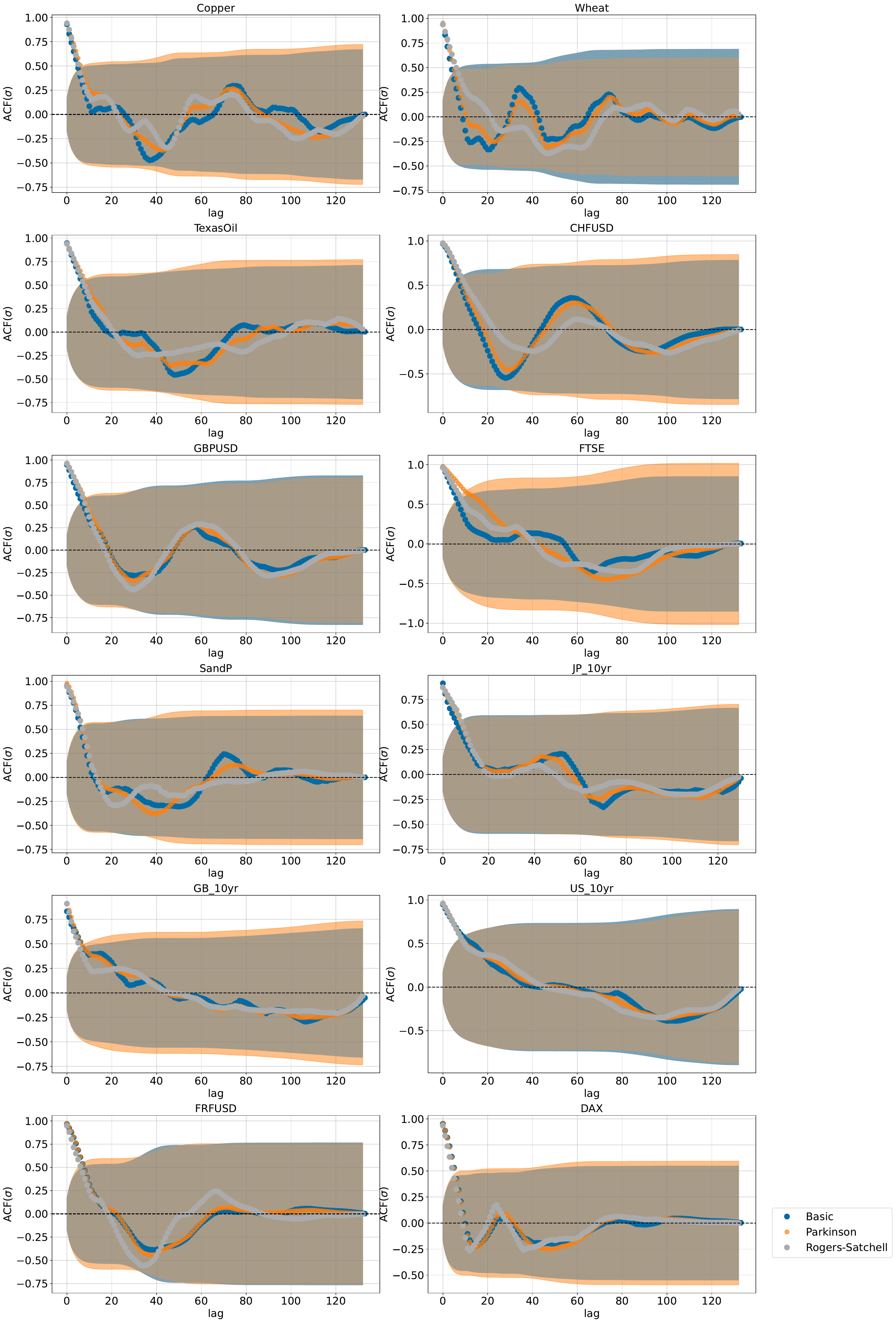}
    \caption{The autocorrelation of the volatility measures described in Section~\ref{sect:volatility} measured for the monthly dataset of Section~\ref{sect:datasets}. The agreement between different metrics are evident. The extent to which this measure remains positive depends on the average size of the volatility cluster specific to each market. The shaded regions are 95\% confidence intervals of the measurements. The time window over which the historical volatility is computed is constant for all the assets to allow comparison and is 12 months.}
    \label{fig:volas_acf_mon}
    \end{figure}

\begin{figure}
    \centering
    \includegraphics[scale=0.22]{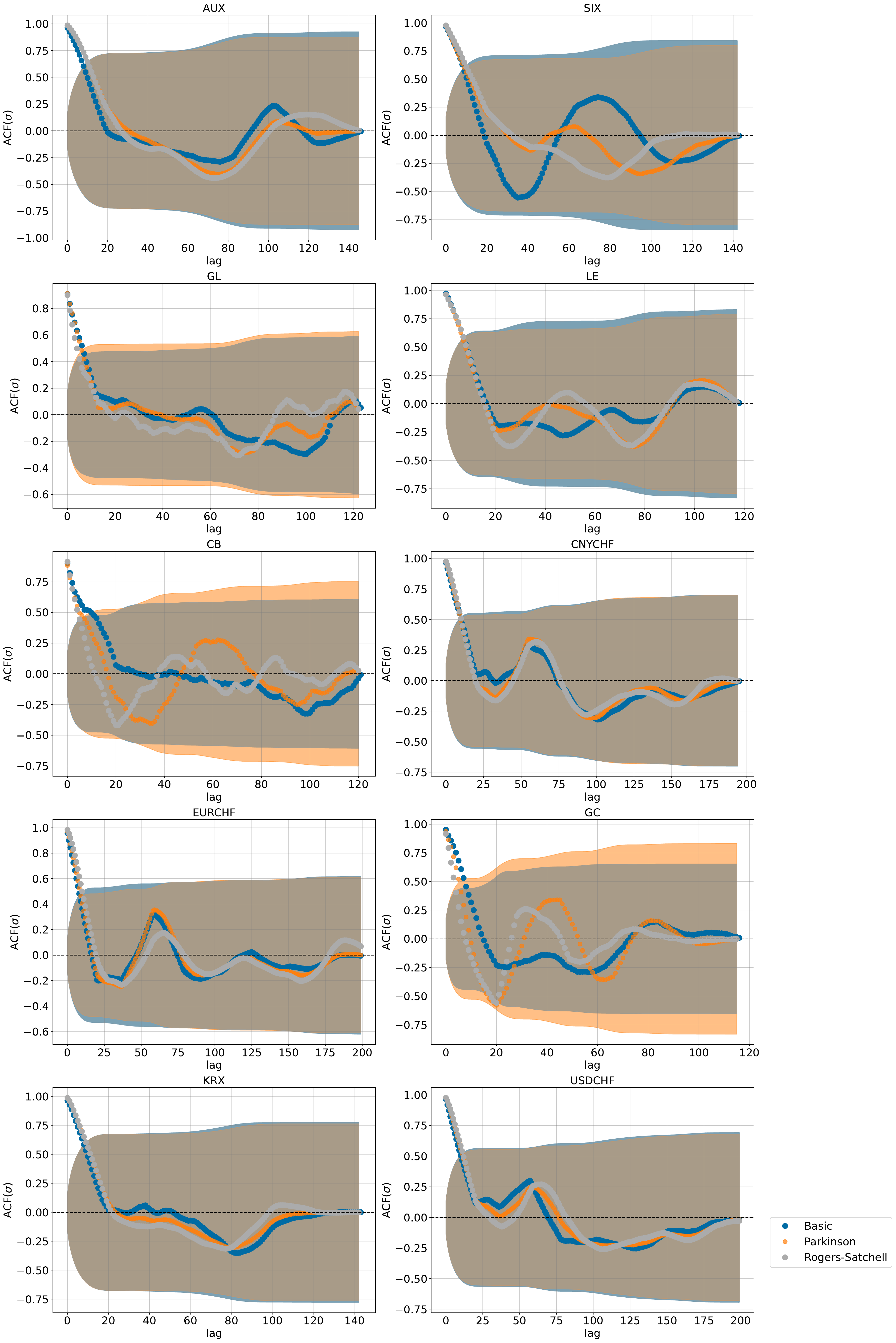}
    \caption{The autocorrelation of the volatility measures described in Section~\ref{sect:volatility} measured for the daily dataset of Section~\ref{sect:datasets}. The agreement between different metrics are evident. The extent to which this measure remains positive depends on the average size of the volatility cluster specific to each market. The shaded regions are 95\% confidence intervals of the measurements. The time window over which the historical volatility is computed is constant for all the assets to allow comparison and is 21 days. }
    \label{fig:volas_acf_day}
    \end{figure}

\begin{figure}
    \centering
    \includegraphics[scale=0.22]{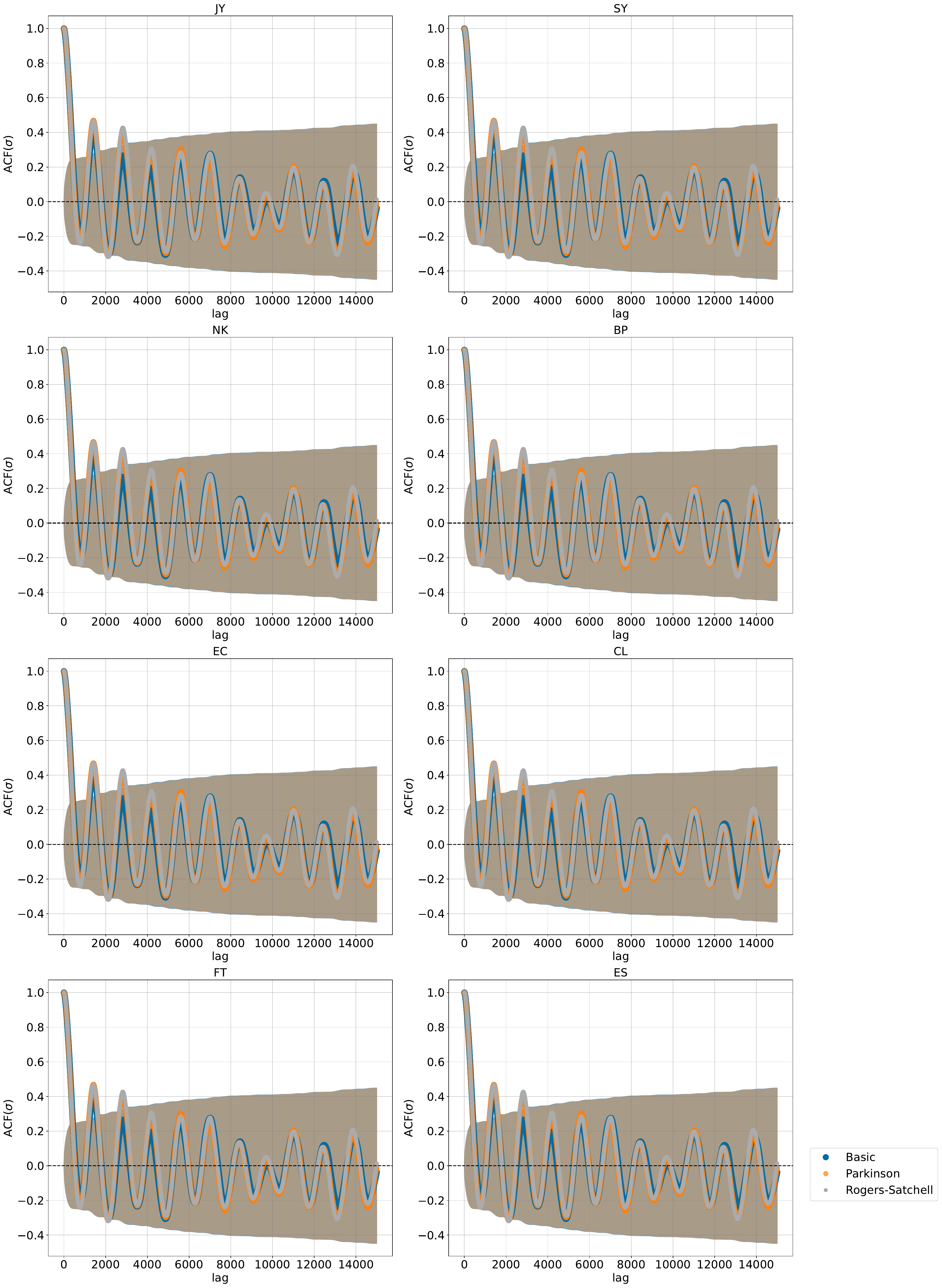}
    \caption{The volatility measures described in Section~\ref{sect:volatility} measured for the high-frequency dataset of Section~\ref{sect:datasets}. The agreement between different metrics are evident. The horizontal axis shows the data point index in the time series and the vertical axis is the measure of volatility. The time window over which the historical volatility is computed is constant for all the assets to allow comparison and is 1 day. One could also see the relative periodicity in the volatility of the high frequency prices induced by scheduled technical underlying events in those markets \cite{Dette2022effect}. Also standard deviation captures the daily fluctuations in closing prices, which can be more volatile than the range between high and low prices used in Parkinson measure or the intra-day movements considered in Rogers-Satchell measure.}
    \label{fig:volas_hf_two}
    \end{figure}

\begin{figure}
    \centering
    \includegraphics[scale=0.22]{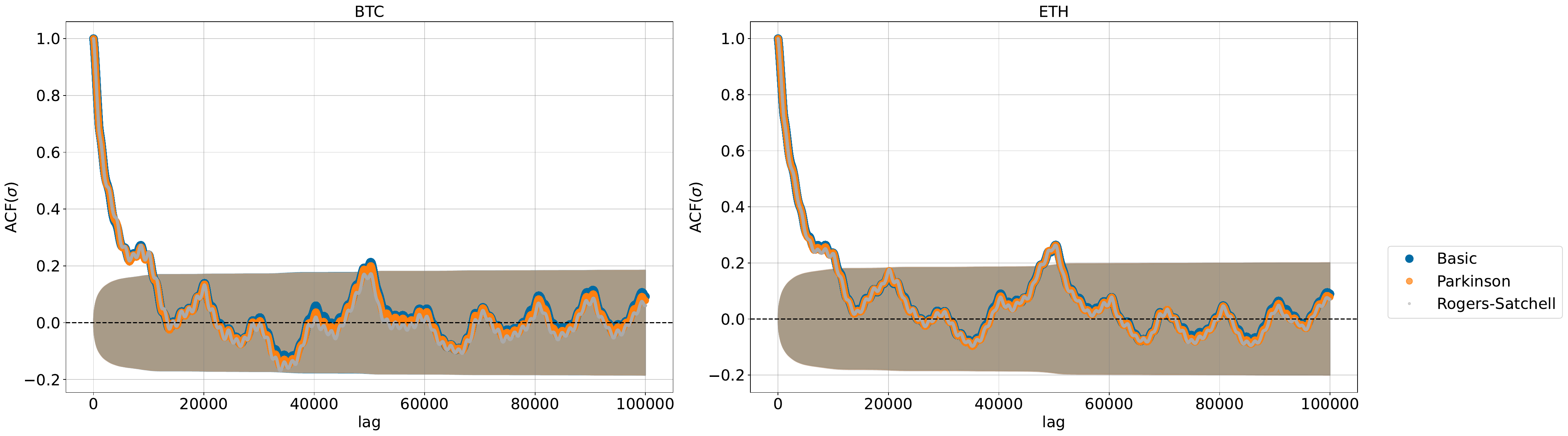}
    \caption{The volatility measures described in Section~\ref{sect:volatility} measured for the cryptocurrency dataset of Section~\ref{sect:datasets}. The agreement between different metrics are evident. The horizontal axis shows the data point index in the time series and the vertical axis is the measure of volatility. The time window over which the historical volatility is computed is constant for all the assets to allow comparison and is 1 day. If compared to the traditional case, one can see in general the volatility decays slower and enters the shaded region slower.}
    \label{fig:volas_acf_cryptos}
    \end{figure}

    Figures~\ref{fig:volas_mon} to~\ref{fig:volas_cryptos} illustrate that various metrics of historical volatility are closely aligned for different assets in both traditional and cryptocurrency markets. This pattern is consistent across different time scales. Consequently, similar behavior in the autocorrelation of volatility is anticipated, regardless of the metric used. The results for these autocorrelations are presented in Figures~\ref{fig:volas_acf_mon} to \ref{fig:volas_acf_cryptos}. 
    We observe both visually and quantitatively a clustering behavior across different markets. The slow decay of realized volatility autocorrelation indicates the speed at which the market transitions from one cluster to another in terms of lag. On a 1-minute time scale, the periodicity is well-reflected in the autocorrelation of volatility (see also \cite{Dette2022effect}). It is noteworthy that measurements may suggest a different speed of this phenomenon in Asian markets compared to others globally. However, confirming this would require further investigation with a larger dataset. 
    We refrain from reporting averaged autocorrelations because the size and intensity of the behavior can vary from market to market. Nonetheless, the clustering effect remains a common feature across different asset classes and should be studied on an asset-by-asset basis.

    \subsection{Leverage effect}\label{sect:leverage-effect}
    \paragraph{Description} This fact asserts that the measures of the volatility of an asset price, such as those discussed in Section~\ref{sect:volatility}, are negatively correlated with past log-returns of that asset. More explicitly, this fact asserts that for small $\delta>0$ and $t\in I$, we have
    \begin{equation*}
        \operatorname{Corr}(r_{t},\sigma_{t+\delta})<0.
    \end{equation*}
    Interestingly, this fact is \enquote{time-asymmetric} since no claim is made that future returns are negatively correlated with past volatility. \cite{Bouchaud_2001, Bouchaud_stylized}  

    \paragraph{Statistical test}
    We compute the cross-correlations between returns and lagged volatility measures.
    
\begin{figure}
    \centering
    \includegraphics[scale=0.25]{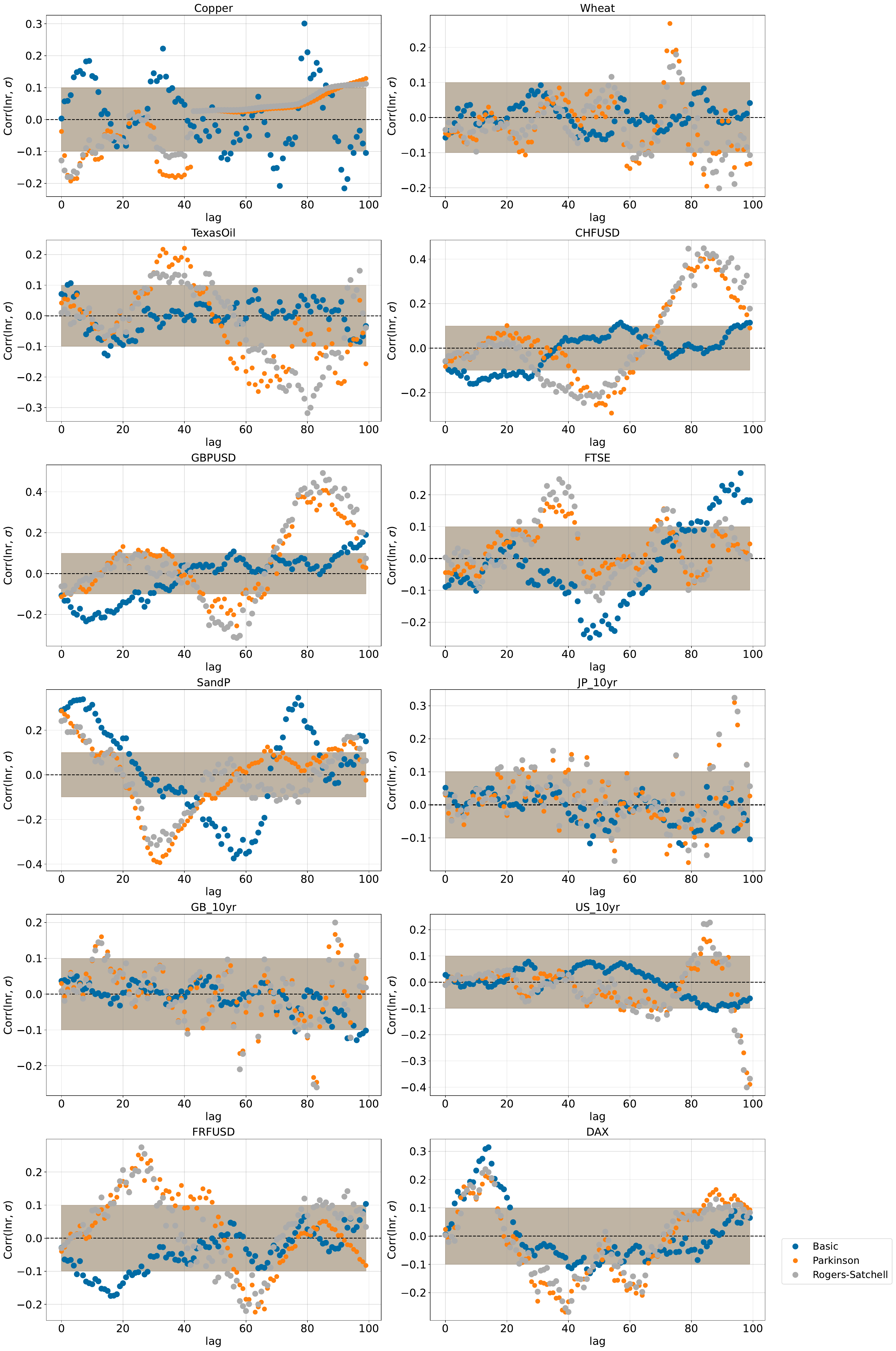}
    \caption{Monthly cross-correlation between historical volatility and log-returns across different time scales and market types. Shaded regions represent 90\% confidence intervals for different volatility measures. The negative correlation at negative lags demonstrates the leverage effect, where past negative returns are correlated with increased future volatility.}
    \label{fig:leverage_m}
    \end{figure}

\begin{figure}
    \centering
    \includegraphics[scale=0.25]{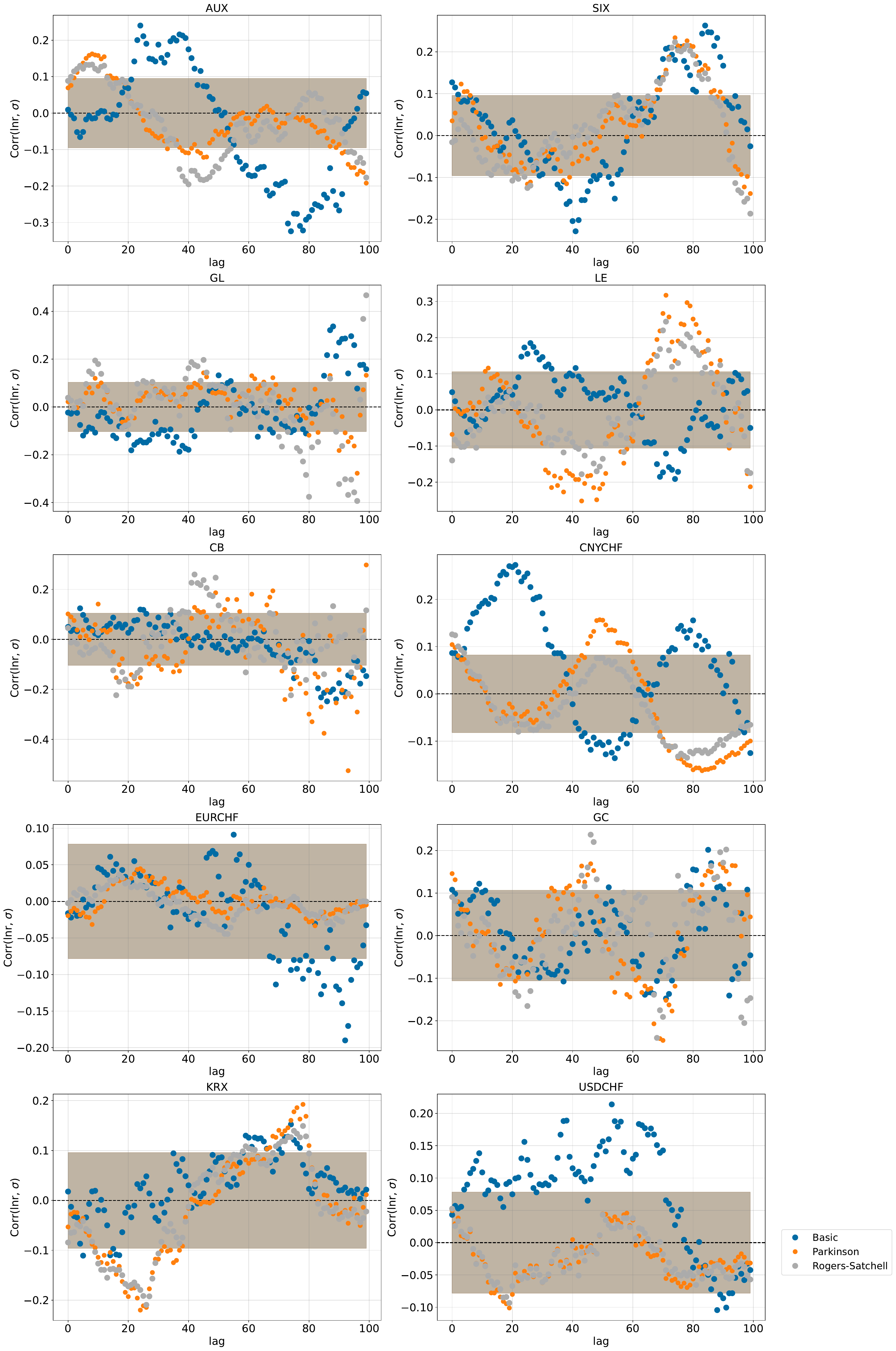}
    \caption{Daily cross-correlation between historical volatility and log-returns across different time scales and market types. Shaded regions represent 90\% confidence intervals for different volatility measures. The negative correlation at negative lags demonstrates the leverage effect, where past negative returns are correlated with increased future volatility.}
    \label{fig:leverage_d}
    \end{figure}

\begin{figure}
    \centering
    \includegraphics[scale=0.25]{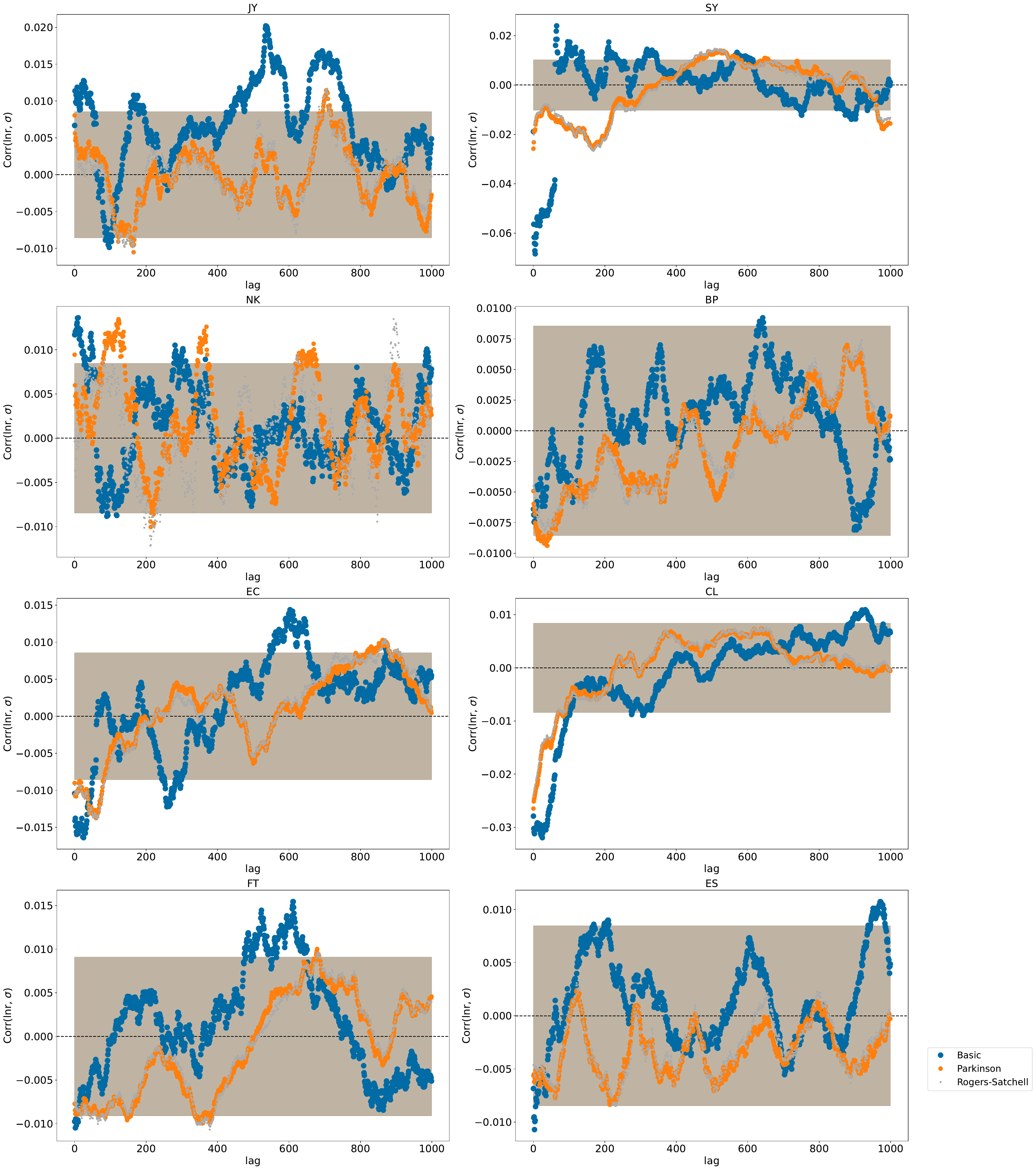}
    \caption{1-minute cross-correlation between historical volatility and log-returns across different time scales and market types. Shaded regions represent 90\% confidence intervals for different volatility measures. The negative correlation at negative lags demonstrates the leverage effect, where past negative returns are correlated with increased future volatility.}
    \label{fig:leverage_hf}
    \end{figure}

\begin{figure}
    \centering
    \includegraphics[scale=0.25]{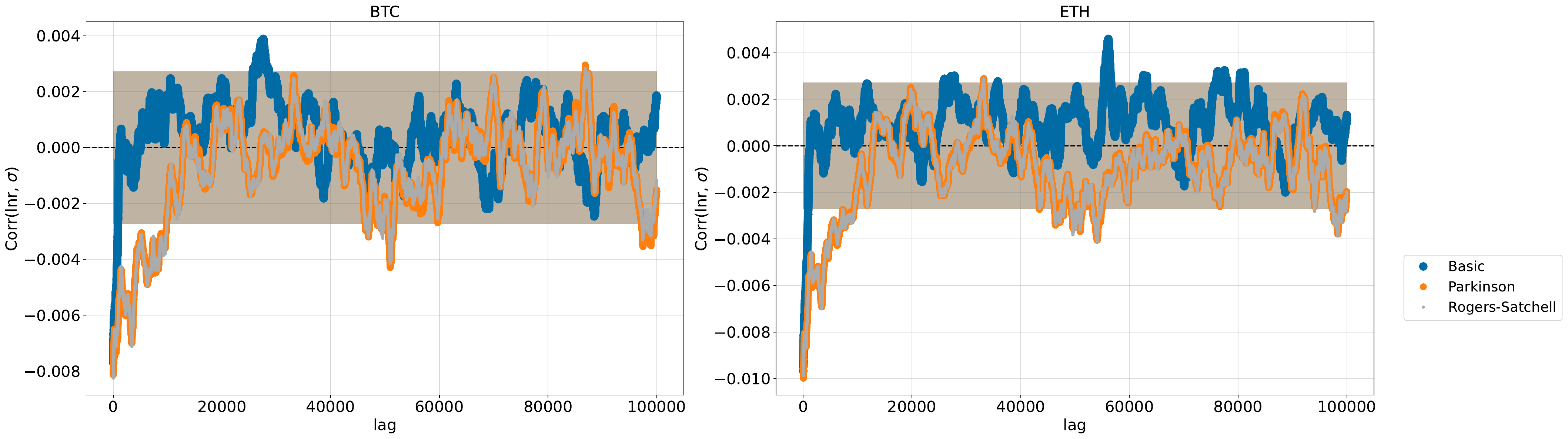}
    \caption{Cryptocurrencies cross-correlation between historical volatility and log-returns across different time scales and market types. Shaded regions represent 90\% confidence intervals for different volatility measures. The negative correlation at negative lags demonstrates the leverage effect, where past negative returns are correlated with increased future volatility.}
    \label{fig:leverage_c}
    \end{figure}

    \paragraph{Result}
    The results are presented in Figures~\ref{fig:leverage_m}, \ref{fig:leverage_d}, \ref{fig:leverage_hf}, and \ref{fig:leverage_c}, which illustrate the relationship between past log-returns and volatility across multiple time scales and market types. These figures show that the sign of the movement is negatively correlated between log-returns and volatilities. This suggests that when a negative return occurs in the market and the magnitude exceeds the average negative magnitudes expected by market participants, a period of high volatility is triggered. Although we hesitate to make a definitive conclusion due to the confidence interval, one can generally observe the negative correlation at various lags. Based on these measurements, the overall conclusion cannot be made that the opposite sign necessarily occurs for every individual asset immediately after a sharp negative change in the log-return. However, for many of the assets in our datasets, this feature holds, as can be seen from the plots.

    \subsection{Volume/volatility correlation}\label{sect:volume-volatility}
    \paragraph{Description} This stylized fact asserts that trading volume is positively correlated with the volatility of an asset. That is, if $V_t$ is used to denote the (random) volume of the asset at time $t$, then
    \begin{equation*}
        \operatorname{Corr}(V_t, \sigma_t)>0.
    \end{equation*}

    \paragraph{Statistical test} Various statistical measures can be used to assess this fact. Here, we propose visualizing a contemporaneous test of the two time series using a scatter plot and quantifying the positivity of the cross-correlation using the Pearson correlation coefficient.
    
\begin{figure}
    \centering
    \includegraphics[scale=0.35]{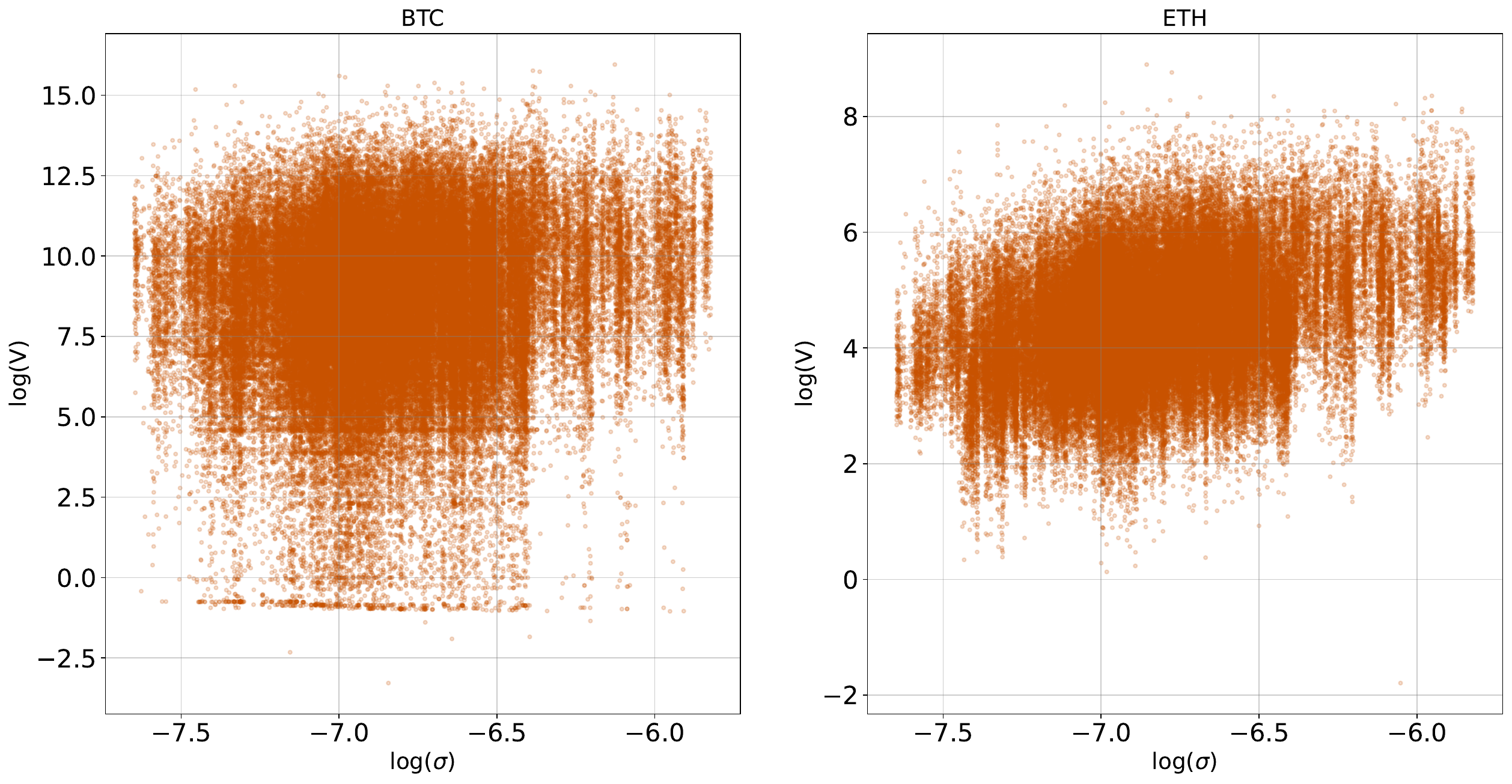}
    \caption{ The scatter plot to demonstrate the positivity of the contemporaneous volatility with volume in cryptocurrency datasets. The Pearson correlations are $r_{BTC}=0.093$, and $r_{ETH}=0.29$}
    \label{fig:vol_vol_c}
    \end{figure}

\begin{figure}
    \centering
    \includegraphics[scale=0.3]{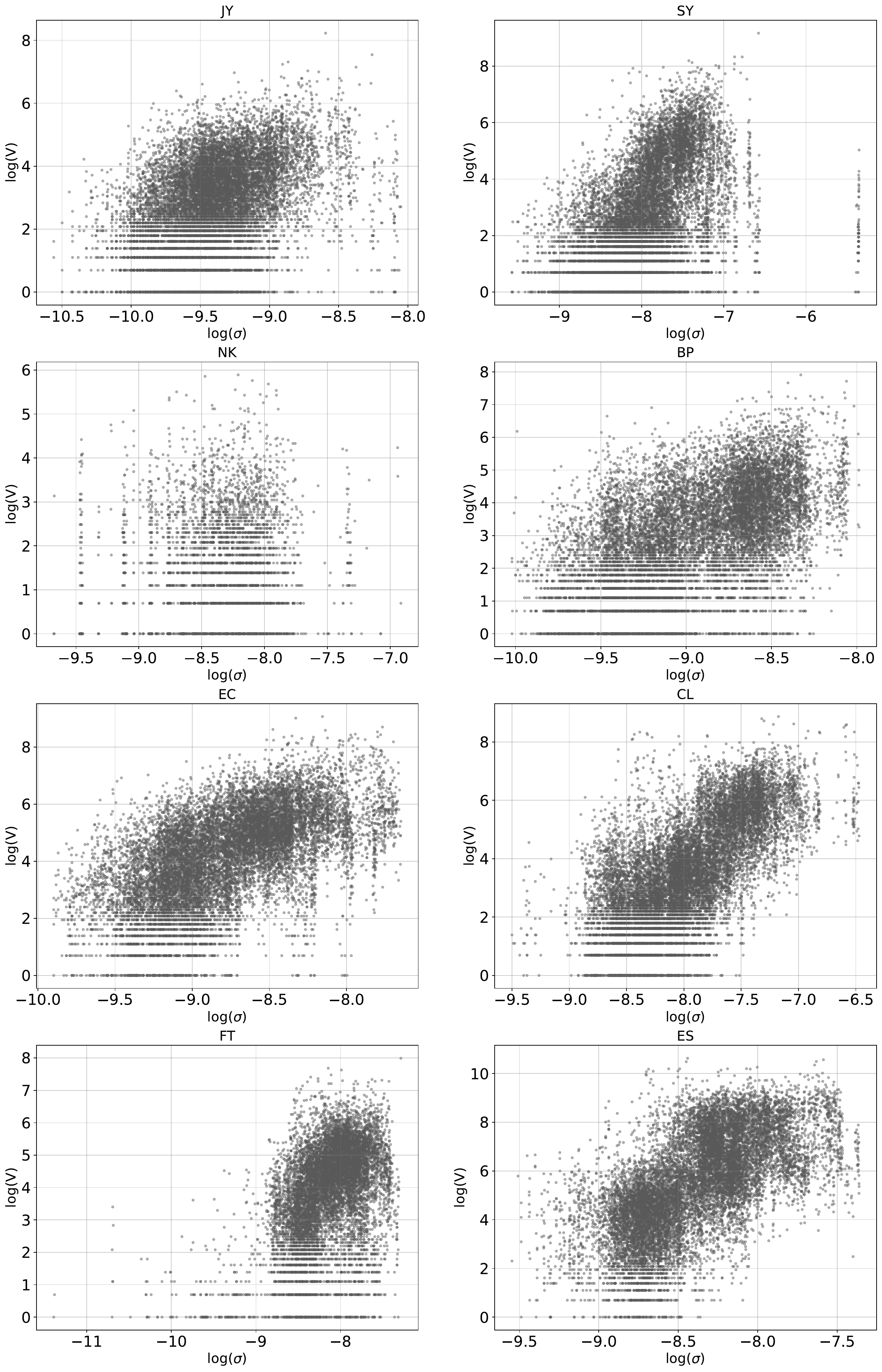}
    \caption{ The scatter plot to demonstrate the positivity of the contemporaneous volatility with volume in 1-minute traditional futures market datasets. The Pearson correlations are $r_{JY}=0.28$, $r_{SY}=0.17$, $r_{NK}=0.059$, $r_{BP}=0.37$, $r_{EC}=0.39$, $r_{CL}=0.47$, $r_{FT}=0.38$, $r_{ES}=0.38$}
    \label{fig:vol_vol_hf}
    \end{figure}

    \paragraph{Result}
    Due to missing values in the traded volume for the daily and monthly datasets, we only report the measurements for the intra-daily datasets in Section~\ref{sect:datasets} for both traditional and cryptocurrency markets. The measurements can be found in Figures~\ref{fig:vol_vol_hf} and~\ref{fig:vol_vol_c}. We confirm that this fact is observed for all assets within the dataset. While the correlation is consistently positive, its strength, as measured by the Pearson coefficient, varies notably across assets, being relatively weak for some markets like Nikkei futures or Bitcoin (see figure captions~\ref{fig:vol_vol_hf} and~\ref{fig:vol_vol_c}).

    \subsection{Unconditional heavy tail}
    \label{sect:uncond}
    \paragraph{Description} This stylized fact asserts that the distribution of $r_{t,\tilde t}$ has a heavy-tailed unconditional distribution on both sides if $\tilde t, t\in I$ with $\tilde t\neq t$.
    
    \begin{definition}[Heavy-tailed distribution]
        A probability distribution $\mu$ on $\R$ is called \emph{heavy-tailed to the right} (respectively, to the left) if, for all $t>0$ (respectively $t<0$), we have $\int_\R \e^{tx}\,\mathrm d\mu(x)=\infty$. It is called \emph{heavy-tailed on both sides} if $\int_\R \e^{tx}\,\mathrm d\mu(x)=\infty$ for all $t\in\R\setminus\set 0$.
    \end{definition}
    
    For conciseness, we present the findings of this section together with those from Sections~\ref{sect:cond} and \ref{sect:gain-loss-asymmetry} in Figure~\ref{fig:tail_alpha}.
    
    \subsection{Conditional heavy tail}\label{sect:conditional_tail}\label{sect:cond}
    \paragraph{Description} Heavy tails in log-returns may arise due to volatility clustering. This stylized fact asserts that even when volatility clustering is accounted for (by conditioning on the volatility), the distribution of returns is still heavy-tailed. More specifically, it asserts that the distribution of $r_{t,\tilde t}$ conditional on $\sigma_t$ is heavy-tailed for $t,\tilde t\in I$ with $t\neq\tilde t$. 

    \paragraph{Statistical test}
     To assess the heavy-tail nature of return distributions, we compute the empirical distribution function (EDF) and measure the exponent with which its tail decays. For the unconditional case, we analyze the raw log-return data, corrected for the historical basic volatility. In the conditional case, we first correct for volatility using a GARCH(1,1) model, which accounts for the autocorrelation in the volatility process, rather than using historical volatility measures. After computing the EDF in each scenario, we focus specifically on the extreme tails by selecting a small fraction of data points in both the negative and positive tails. We then fit a line in log-log scale to estimate the tail exponent $\alpha$ in the power-law relationship 
     
     \begin{equation*}
        P(X>x)\propto x^{-\alpha},
        \label{eq:edf}
     \end{equation*}
     
     which characterizes the rate of decay in the probability distribution tail.

    \paragraph{Result}
    Before looking at quantitative estimates, Figures~\ref{fig:agg_gaus_m}, \ref{fig:agg_gaus_d}, \ref{fig:agg_gaus_hf} and \ref{fig:agg_gaus_cryptos} already show that the unconditional tail distribution of the log-returns is heavier than a Gaussian. 
    The results of our conditional and unconditional heavy tail analysis are presented in Figure~\ref{fig:tail_alpha}, alongside the findings from Section~\ref{sect:gain-loss-asymmetry}. We analyze these results together due to their interconnected nature. As shown in the figure, both the unconditional and conditional distributions exhibit heavy tails, with the conditional distribution (after GARCH(1,1) correction) showing less extreme but still heavy-tail behavior. Additionally, we observe a consistent pattern of heavier left tails compared to right tails across market types and time scales, providing evidence for both the conditional heavy tail property and gain/loss asymmetry discussed in Section~\ref{sect:gain-loss-asymmetry}.
    
    \subsection{Gain/Loss asymmetry}
    \label{sect:gain-loss-asymmetry}
    \paragraph{Description} This stylized fact asserts that (for non-currencies), the distribution of log returns has a heavier left-tail than right-tail.

    \paragraph{Statistical test}
    We estimate the tail exponents for both the left and right tails of the return distributions. We check for a systematic difference between these exponents, specifically a smaller exponent for the left tail (indicating heavier tails) compared to the right tail.
    
\begin{figure}
    \centering
    \includegraphics[scale=0.3]{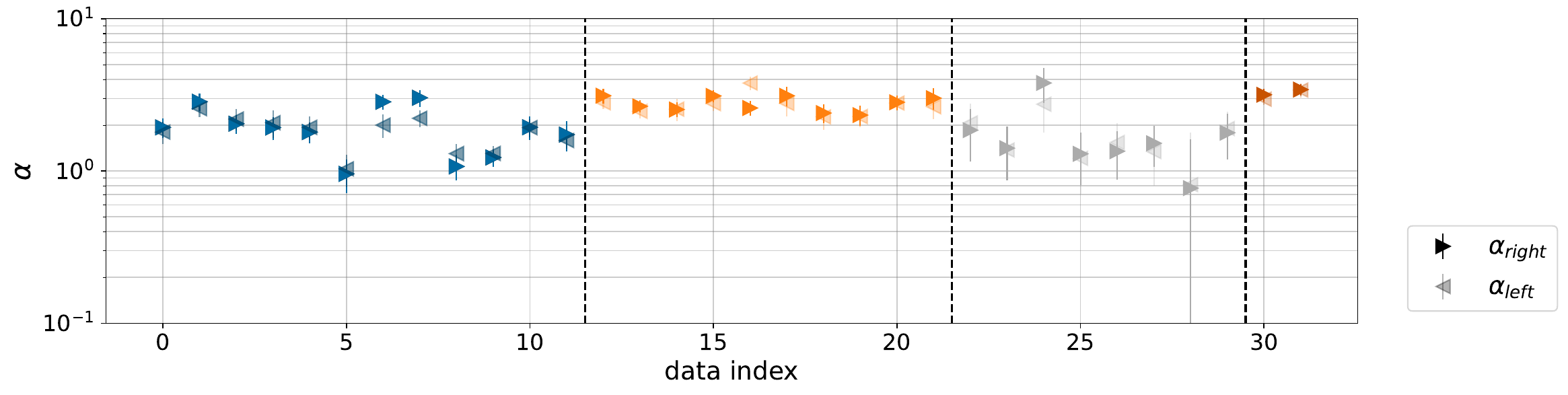}
    \includegraphics[scale=0.3]{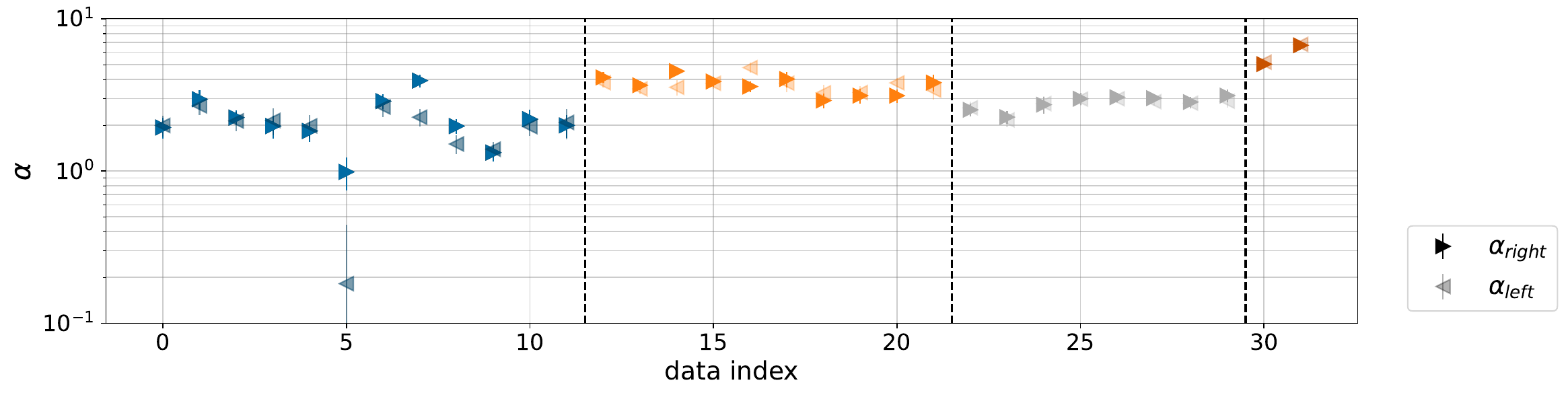}
    \caption{Right and left tail exponents for all markets analyzed in this study. Right-pointing arrows represent right tail indices, while left-pointing arrows represent left tail indices. The upper panel shows unconditional indices, and the lower panel shows conditionally corrected indices after GARCH(1,1) adjustment. Vertical dashed lines separate different time scale regimes. The horizontal axis in both plots indicates the index number of each time series in the database while the vertical aaxes show $\alpha$ from Equation~\eqref{eq:edf}.}
    \label{fig:tail_alpha}
    \end{figure}

    \paragraph{Result}
    Figure~\ref{fig:tail_alpha} displays the tail exponents for both the unconditional distribution (upper panel) and the conditional distribution after GARCH(1,1) correction (lower panel). Notably, across market types and time scales, the left tail exponents are consistently smaller than the right tail exponents, indicating heavier left tails. This systematic asymmetry provides evidence for the gain/loss asymmetry stylized fact.
    
    \subsection{Aggregational Gaussianity}\label{sect:aggregational-gaussianity}
    \paragraph{Description} This stylized fact in the strong formulation asserts that for $t\in I$ fixed and as $\delta\to\infty$, we have convergence of the 
    \begin{equation}\label{eq:normalized-log-returns}
        \frac{r_{t,t+\delta} - \E(r_{t,t+\delta})}{\sqrt{\V(r_{t,t+\delta})}}
    \end{equation}
    in distribution towards a standard normal distribution. 

    \paragraph{Statistical test}
    We perform the Kolmogorov--Smirnov test, which measures the distance of the empirical distribution of equation~\ref{eq:normalized-log-returns} to a standard normal distribution as well as Anderson-Darling to accommodate a higher sensitivity to the tail events compared to the former. The closer the two distributions are, the smaller this statistic. We also use the Anderson--Darling test, which is more sensitive to tail behavior.

\begin{figure}
    \centering
    \includegraphics[scale=0.3]{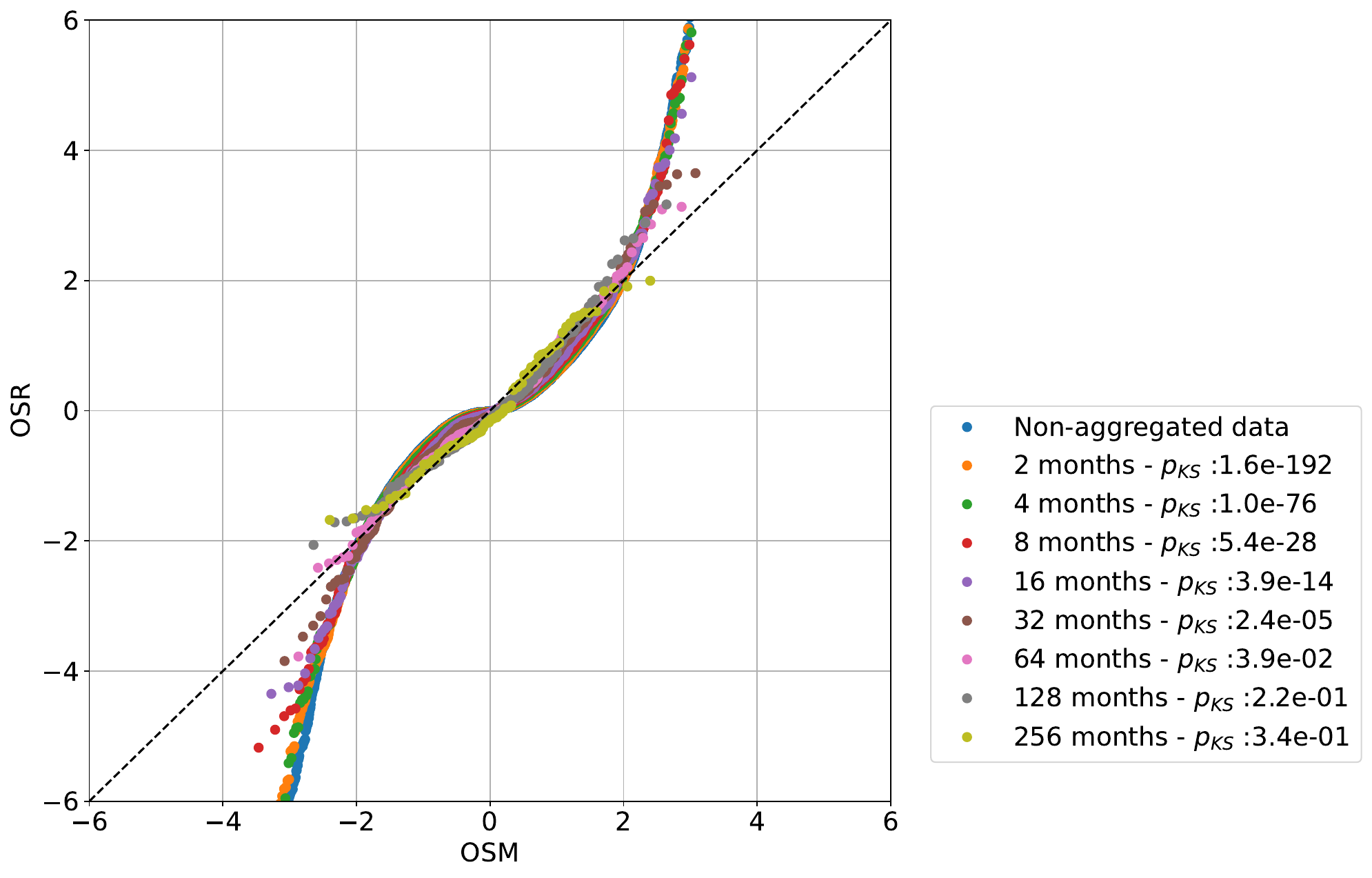}
    \caption{Quantile-quantile plot for monthly data, comparing them to a gaussian with the same mean and variance. KS p-value is shown for comparison.}
    \label{fig:agg_gaus_m}
    \end{figure}

\begin{figure}
    \centering
    \includegraphics[scale=0.3]{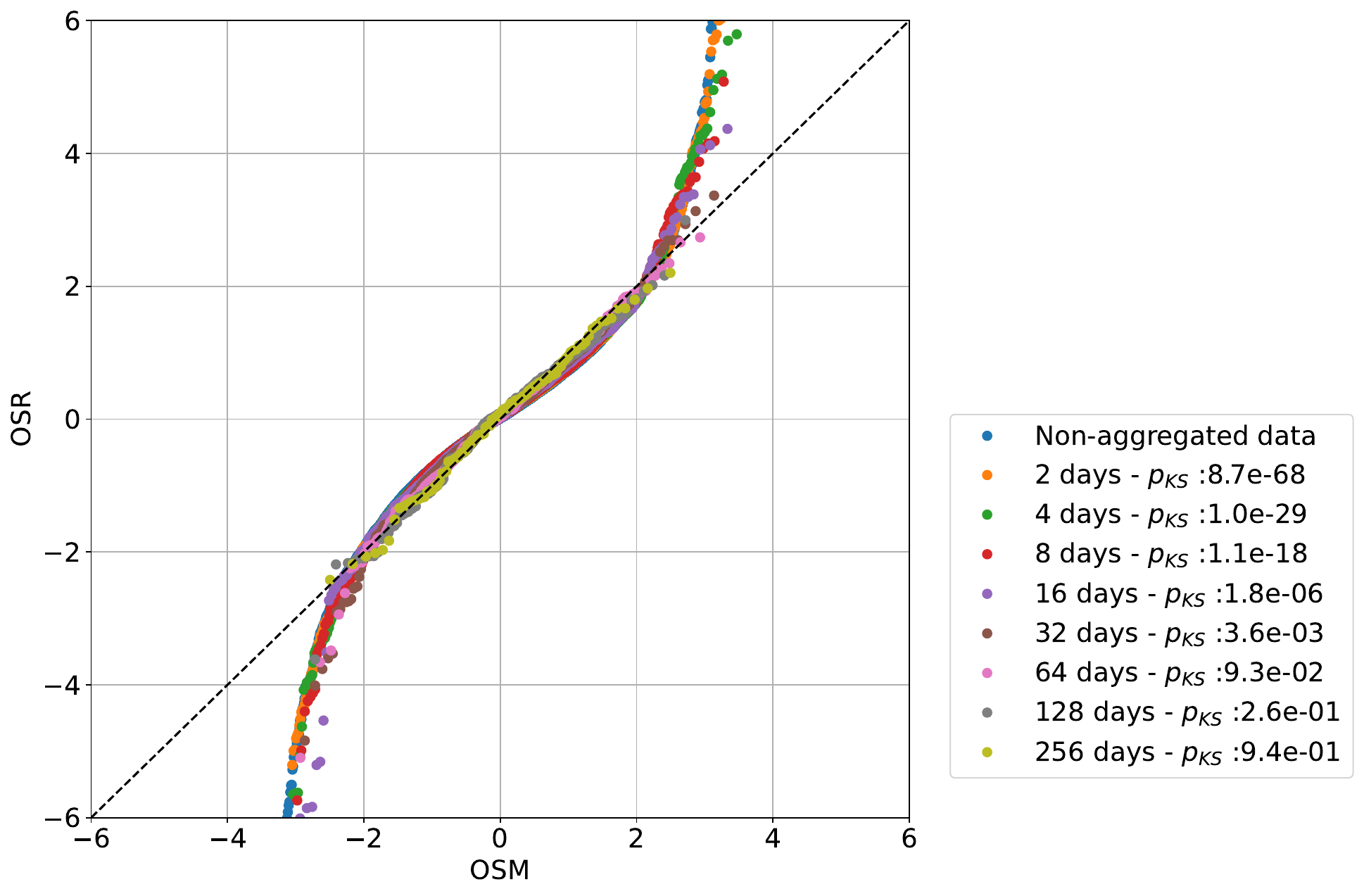}
    \caption{Quantile-quantile plot for daily data, comparing them to a gaussian with the same mean and variance. KS p-value is shown for comparison.}
    \label{fig:agg_gaus_d}
    \end{figure}

\begin{figure}
    \centering
    \includegraphics[scale=0.3]{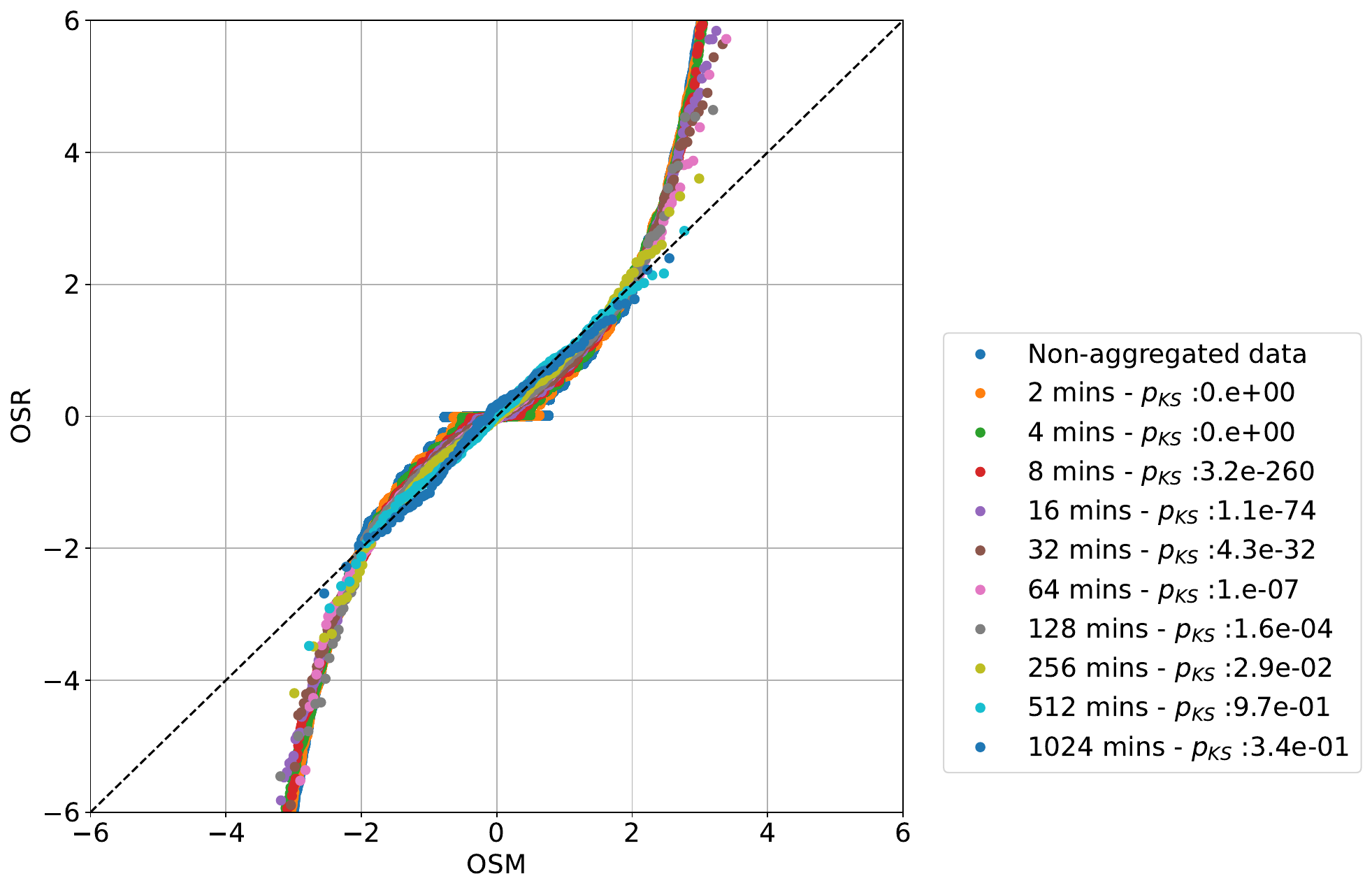}
    \caption{Quantile-quantile plot for 1-minute traditional data comparing them to a gaussian with the same mean and variance. KS p-value is shown for comparison.}
    \label{fig:agg_gaus_hf}
    \end{figure}

\begin{figure}
    \centering
    \includegraphics[scale=0.3]{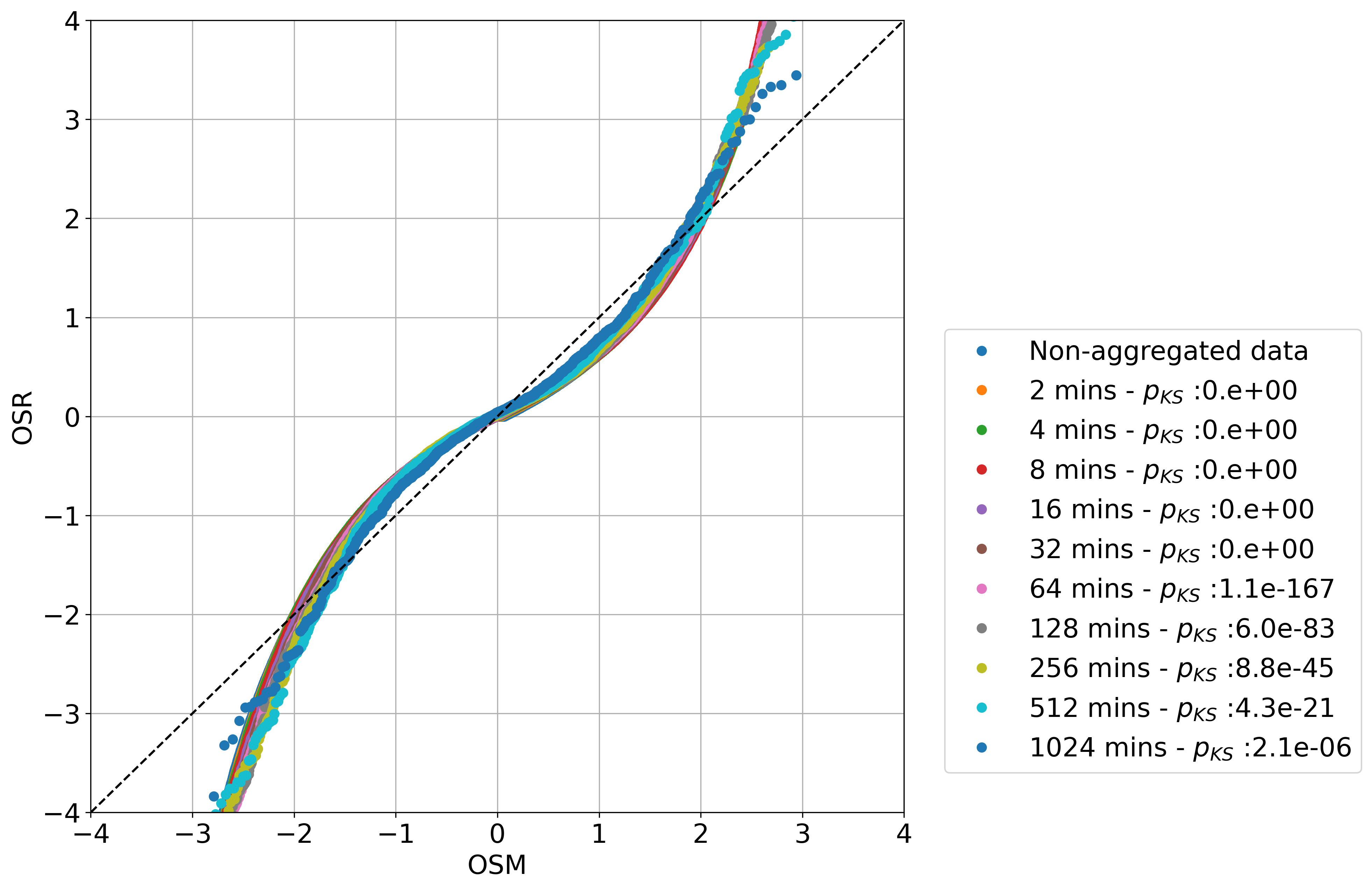}
    \caption{Quantile-quantile plot for cryptocurrency data, comparing them to a gaussian with the same mean and variance. KS p-value is shown for comparison. }
    \label{fig:agg_gaus_cryptos}
    \end{figure}

    \paragraph{Result} 
    If log-returns were independent and identically distributed (iid) with finite variance, the Central Limit Theorem would imply that longer-term log-returns converge to a normal distribution. However, financial returns typically violate these assumptions due to volatility clustering and heavy tails. Nevertheless, figures \ref{fig:agg_gaus_m} to \ref{fig:agg_gaus_cryptos} indicate that as we increase the time intervals over which log-returns are measured, their unconditional distribution does approach normality. This phenomenon may occur because the heavy tails become less pronounced when examining fewer long-term log-returns. 
    
    The Anderson-Darling test results, presented in Table~\ref{tab:anderson_darling}, support the observation of aggregational Gaussianity. However, even if the true distribution does not approach a Gaussian, observing fewer returns over longer time horizons means we are less likely to observe tail events, making the distribution appear more Gaussian simply because we are only seeing the bulk of the distribution. 

\begin{table}
\centering
\begin{tabular}{ |p{3.5cm}||p{1.5cm}|p{1.5cm}|p{1.5cm}|p{1.5cm}|  }\hline
 	Significance levels & $10\%$ &  $ 5\%$ & $1\%$ & AD Statistic \\ \hline
    \multicolumn{5}{|l|}{Monthly} \\\hline
	2 months &  0.656 & 0.787 &  1.092 &  489.3\\  
	8 months &  0.655 & 0.786 & 1.090 & 73.1 \\	
    32 months & 0.652 &0.782 & 1.085 & 8.88 \\	
    128 months & 0.641& 0.769 & 1.068 & 1.45 \\
    256 months &  0.628& 0.754& 1.046 & 0.81\\ \hline
    \multicolumn{5}{|l|}{Days} \\\hline
	2 days & 0.656 & 0.787& 1.092&  190.9\\  
    32 days &  0.653 & 0.783 & 1.087& 4.68 \\	
    64 days & 0.650 & 0.78 & 1.082 & 1.03 \\
    256 days &   0.634 &  0.761 & 1.056 & 0.26\\ \hline
    \multicolumn{5}{|l|}{Minutes} \\\hline
	2 minutes &  0.656 & 0.787 & 1.092 & 4064.6\\  
    16 minutes  &0.656 & 0.787 & 1.091 & 175.8 \\	
    32 minutes  & 0.655& 0.786 & 1.091 & 72.0 \\
    64 minutes  & 0.655 &0.785 &1.09 & 19.6\\ 
    512 minutes  & 0.646& 0.775 & 1.075 & 0.24\\ 
    1024 minutes  & 0.637& 0.764 & 1.06 & 0.68 \\ \hline
    \multicolumn{5}{|l|}{Cryptocurrencies} \\\hline
	2 minutes &  0.656& 0.787 & 1.092 & 16580.3\\  
    16 minutes  &0.656 & 0.787 & 1.092 & 1899.3 \\	
    32 minutes  & 0.656& 0.787 & 1.092 & 979.0 \\
    64 minutes  & 0.656 &0.787 &1.092 & 477.3\\ 
    512 minutes  &0.655& 0.785 & 1.09 & 45.7\\ 
    1024 minutes  & 0.653& 0.784 & 1.088 & 16.6 \\ \hline    
\end{tabular}
\label{tab:anderson_darling}
\caption{The table shows the Anderson-Darling (AD) statistics along with the significance level. Similar to the results of the Kolmogorov-Smirnov (KS) test we confirm the convergence of the distribution to normal, considering the AD test is adjusted to be more sensitive to tails compared to KS.}
\end{table}

    \subsection{Time scale asymmetry}
    \label{sect:ts_asym}

    \paragraph{Description} Time scale asymmetry occurs if the instantaneous log-return series $(r_t)_{t\in I}$ is not invariant under time reversal.

    One example of this time invariance is the leverage effect discussed in section \ref{sect:leverage-effect}, according to which  \emph{past} returns are negatively correlated with \emph{future} volatilities, but no claims about the correlation between future returns and past volatilities is made, see e.g. \cite{Bouchaud_2001}.  

    Empirically, another effect, named the \emph{Zumbach effect} \cite{Zumbach_2001}, has been observed. A stochastic model for market data possesses the Zumbach effect if, for a sufficiently small time lag $\delta>0$ and $t\in I$, we have 
    \begin{equation}\label{eq:Zumbach-effect}
    \E\left(r_t^2\sigma_{t+\delta}^2\right)>\E\left(r_{t+\delta}^2\sigma_t^2\right).
    \end{equation}

    \paragraph{Statistical test}
    In order to test whether \eqref{eq:Zumbach-effect} is present in empirical data, we use the following test statistic, suggested in \cite[Formula (6b)]{Chicheportiche_2014} (see also \cite{Muller1997,info_flow}):
    \begin{equation*}
        Z(\delta, I)\define \tilde\kC^{(2)}(\delta, I) - \tilde\kC^{(2)}(-\delta, I),
    \end{equation*}
    where, once again, $\delta>0$ denotes the time lag and $I$ denotes the time interval for which we have access to the log-returns and the estimators of volatility as described in Section~\ref{sect:volatility}. We furthermore defined 
    
    \begin{equation*}
        \tilde\kC^{(2)}(\delta, I)\define\sum_{t\in I} (\hat\sigma_t^2-\sum_{\tilde t\in I}\hat\sigma_{\tilde t}^2) r_{t-\delta}^2.
    \end{equation*}

\begin{figure}
    \centering
    \includegraphics[scale=0.25]{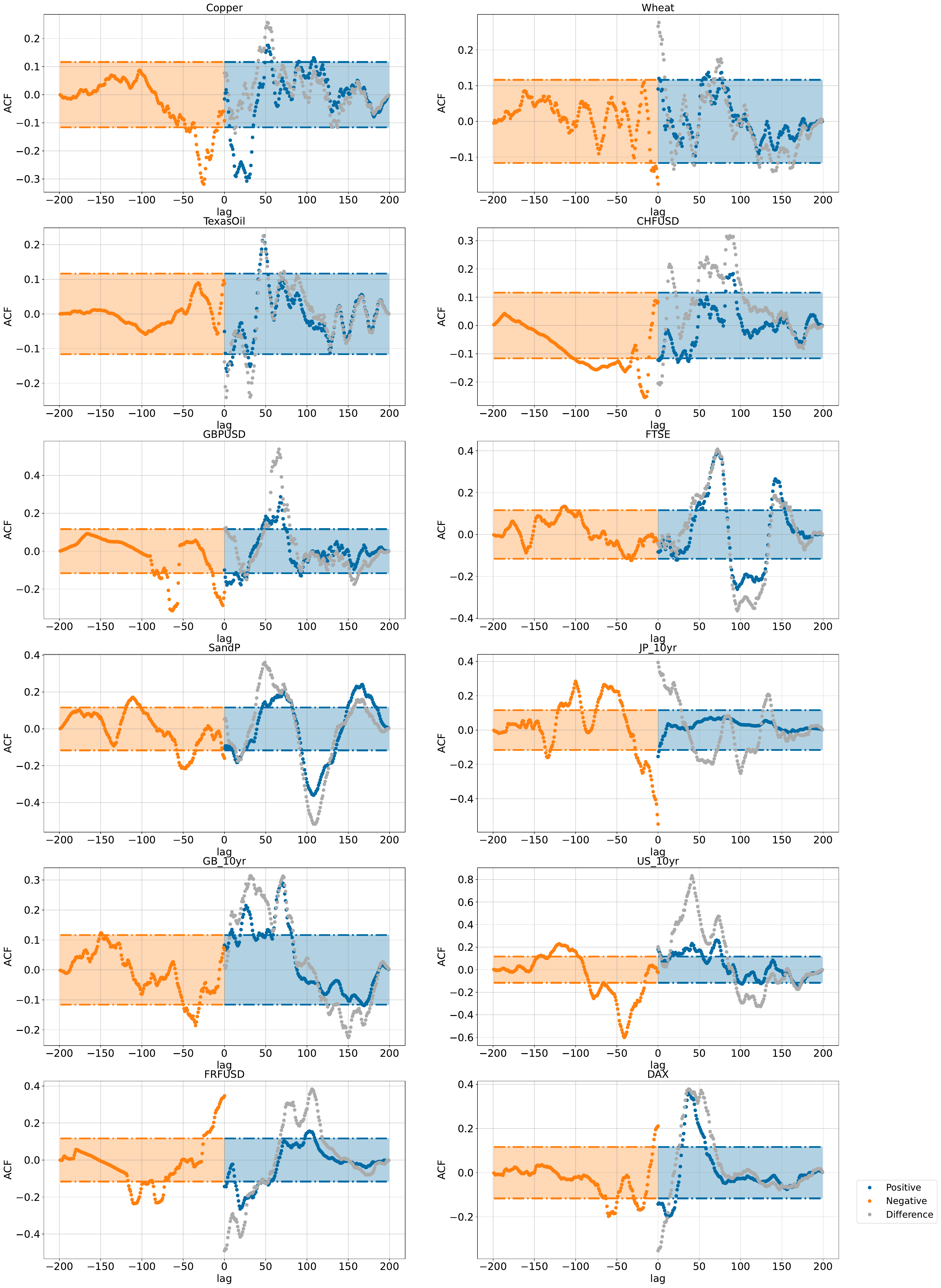}
    \caption{Information flow for the monthly dataset. The confidence interval of 95\% is considered. Gray dots show the difference between the two cross-correlations that differ from being zero beyond the confidence interval.}
    \label{fig:info_flow_m}
    \end{figure}

\begin{figure}
    \centering
    \includegraphics[scale=0.25]{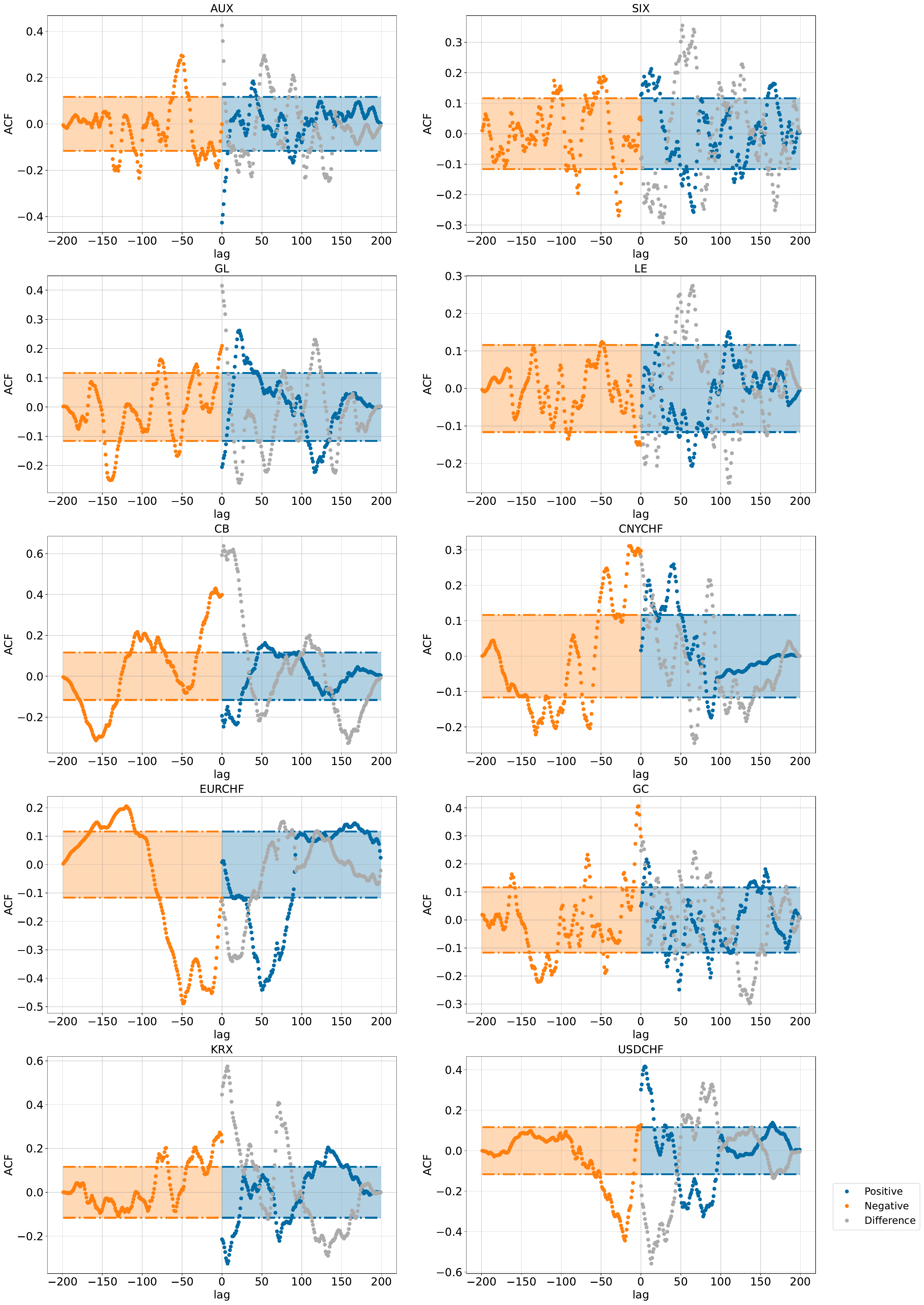}
    \caption{Information flow for the daily dataset. The confidence interval of 95\% is considered. Gray dots show the difference between the two cross-correlations that differ from being zero.}
    \label{fig:info_flow_d}
    \end{figure}

\begin{figure}
    \centering
    \includegraphics[scale=0.25]{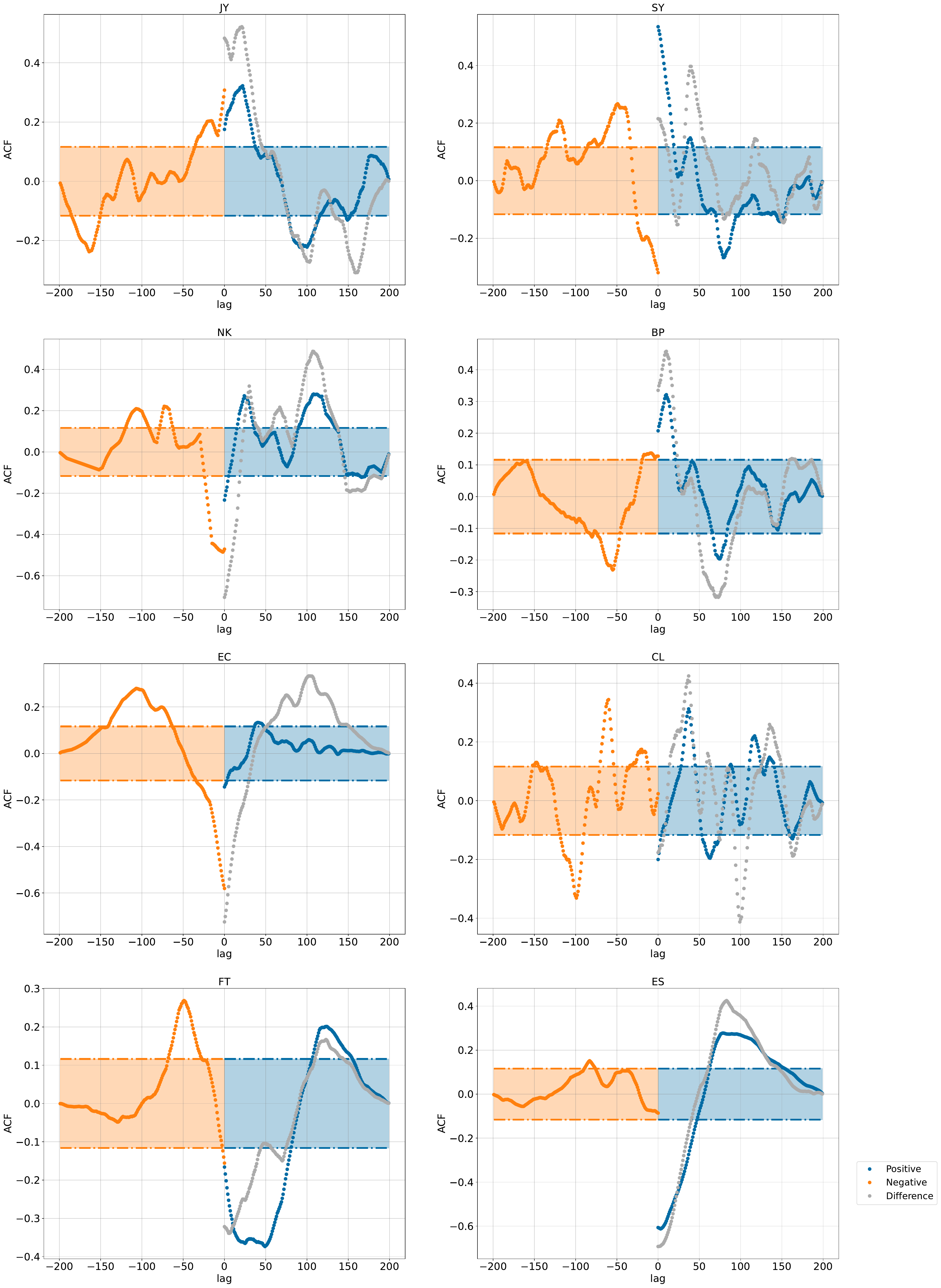}
    \caption{Information flow for the 1-minute futures dataset. The confidence interval of 95\% is considered. Gray dots show the difference between the two cross-correlations that differ from being zero.}
    \label{fig:info_flow_hf}
    \end{figure}

\begin{figure}
    \centering
    \includegraphics[scale=0.3]{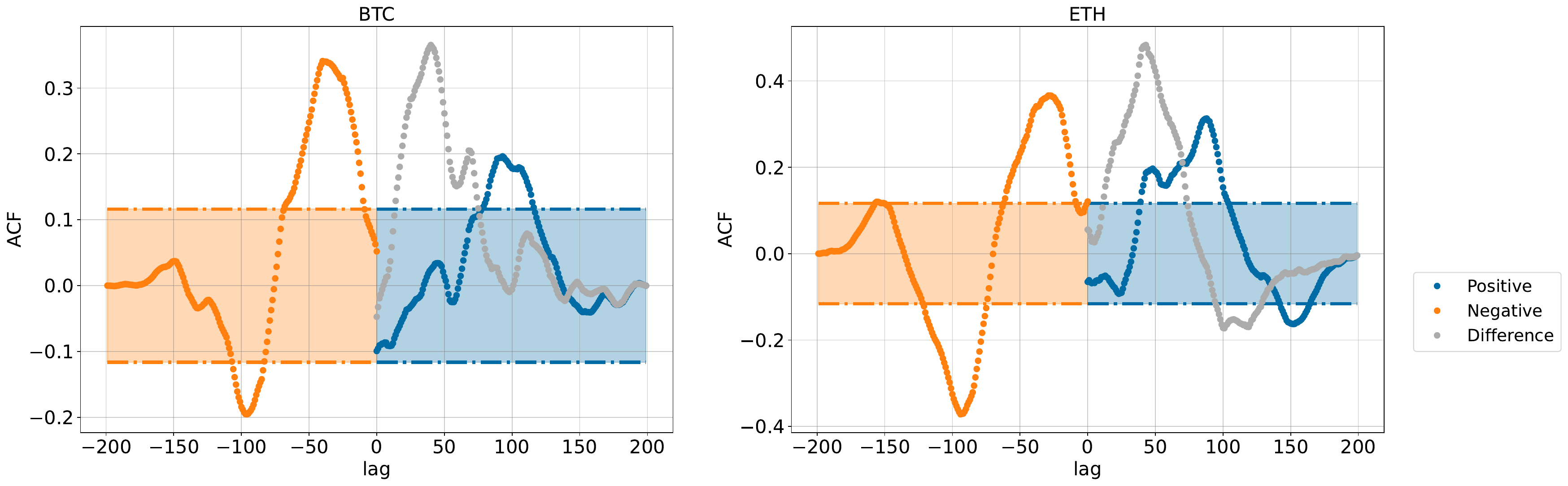}
    \caption{Information flow for the cryptocurrencies dataset. The confidence interval of 95\% is considered. Gray dots show the difference between the two cross-correlations that differ from being zero.}
    \label{fig:info_flow_c}
    \end{figure}

    \paragraph{Result}
    Here we present the result of this test over different assets shown in figures~\ref{fig:info_flow_m} to ~\ref{fig:info_flow_c}. In all the plots we can see that the difference of the cross-correlation measured with different granularity is far from zero within the confidence interval of 95\%. As also discussed in \cite{info_flow,Muller1997,comments} this shows that the information flows across different granularity of markets are asymmetric.

    \section{Final remarks}
    In summary we confirm that we have found evidence in favor of most of the suggested facts proposed by \cite{Cont2001}. We also need to mention that these facts suggested by the article are not universally accepted point-by-point throughout the community, which necessitates ongoing investigation. 
    
    \begin{table}[H]
    \centering
    \caption{Summary of Evidence for Stylized Facts}
    \label{tab:summary_facts}
    \begin{tabularx}{\textwidth}{|X|X|} 
    \hline
    \textbf{Stylized Fact} & \textbf{Evidence Found} \\
    \hline
    1. Absence of autocorrelations (log-returns) (Section~\ref{sect:absence-of-autocorrelations}) & Yes for traditional markets; No for cryptocurrencies (slower decay). \\
    2. Slow decay of autocorrelations (absolute log-returns) (Section~\ref{sect:slow-decay-of-autocorrelations}) & Yes, power-law decay observed, generally consistent with expected exponents. \\
    3. Intermittency (Section~\ref{sect:intermittency}) & Proposed comparisons to benchmarks yield support for intermittency, however, no standard quantitative tests are available. \\ 
    4. Volatility clustering (Section~\ref{sect:volatility-clustering}) & Yes, observed across different markets, time scales, and volatility metrics. \\
    5. Leverage effect (negative correlation between return and future volatility) (Section~\ref{sect:leverage-effect}) & Generally yes, observed negative correlation, though not uniformly immediate across all individual assets. \\
    6. Volume/volatility correlation (Section~\ref{sect:volume-volatility}) & Yes, positive correlation observed for intraday data where volume was available. \\
    7. Conditional heavy tails (Section~\ref{sect:cond}) & Yes, tails remain heavy even after GARCH(1,1) correction. \\
    8. Unconditional heavy tails (Section~\ref{sect:uncond}) & Yes, distributions clearly heavier-tailed than Gaussian. \\
    9. Gain/loss asymmetry (heavier left tail) (Section~\ref{sect:gain-loss-asymmetry}) & Yes, left tails consistently heavier than right tails across markets and conditions. \\
    10. Aggregational Gaussianity (Section~\ref{sect:aggregational-gaussianity}) & Yes, distributions approach normality over longer aggregation intervals, though this apparent convergence may be influenced by observing fewer tail events. \\
    11. Time scale asymmetry (Zumbach effect) (Section~\ref{sect:ts_asym}) & Yes, Zumbach effect indicative of asymmetric information flow observed. \\
    \hline
    \end{tabularx}
    \end{table}
    
    Table~\ref{tab:summary_facts} summarizes our findings regarding the eleven stylized facts discussed.

    As shown in Table~\ref{tab:summary_facts}, our analysis confirms most stylized facts across the datasets, with key differences in cryptocurrency autocorrelation and the universality of the leverage effect. These patterns are valuable for validating financial models, but cryptocurrencies need special attention due to their unique market structure. Some facts, like the cross anti-correlation of volatility and returns, may represent averaged behaviors rather than being applicable to individual assets. Future research may repeat our approach in this paper with larger and more diverse datasets.
    
\section*{Acknowledgments}
The authors would like to thank Ashkan Nikeghbali and Christof Schmidhuber for suggesting and supporting the project, as well as Rama Cont and Marc Wildi for useful discussions. This research is supported by grant no. PT00P2 206333 from the Swiss National Science Foundation.  
\appendix

\printbibliography    

@article{Cont2001,
  author = {Cont, Rama},
  issn = {1469-7688},
  journal = {Quantitative Finance},
  number = 2,
  pages = {223--236},
  publisher = {Routledge},
  title = {Empirical properties of asset returns: stylized facts and statistical
	issues},
  url = {http://www.informaworld.com/10.1080/713665670},
  volume = 1,
  year = 2001
}

@InProceedings{Cont_Beyond,
author="Cont, Rama
and Potters, Marc
and Bouchaud, Jean-Philippe",
editor="Dubrulle, Bérengère
and Graner, François
and Sornette, Didier",
title="Scaling in Stock Market Data: Stable Laws and Beyond",
booktitle="Scale Invariance and Beyond",
year="1997",
publisher="Springer Berlin Heidelberg",
address="Berlin, Heidelberg",
pages="75--85",
}

@InProceedings{Cont_longRange,
author="Cont, Rama",
editor="L{\'e}vy-V{\'e}hel, Jacques
and Lutton, Evelyne",
title="Long range dependence in financial markets",
booktitle="Fractals in Engineering",
year="2005",
publisher="Springer London",
address="London",
pages="159--179",
}

@article{Chakraborti2009,
author = {Anirban Chakraborti, Ioane Muni Toke, Marco Patriarca and Frédéric Abergel},
title = {Econophysics review: I. Empirical facts},
journal = {Quantitative Finance},
volume = {11},
number = {7},
pages = {991--1012},
year = {2011},
publisher = {Routledge},
doi = {10.1080/14697688.2010.539248},
URL = {https://doi.org/10.1080/14697688.2010.539248},
eprint = {https://doi.org/10.1080/14697688.2010.539248}
}

@article{Katahira_spec_game,
title = {Development of an agent-based speculation game for higher reproducibility of financial stylized facts},
journal = {Physica A: Statistical Mechanics and its Applications},
volume = {524},
pages = {503-518},
year = {2019},
issn = {0378-4371},
doi = {https://doi.org/10.1016/j.physa.2019.04.157},
url = {https://www.sciencedirect.com/science/article/pii/S0378437119305333},
author = {Kei Katahira and Yu Chen and Gaku Hashimoto and Hiroshi Okuda},
}

@INPROCEEDINGS{Stylized_modern,
  author={Ratliff-Crain, Ethan and Oort, Colin M. Van and Bagrow, James and Koehler, Matthew T. K. and Tivnan, Brian F.},
  booktitle={2023 IEEE International Conference on Big Data (BigData)}, 
  title={Revisiting Stylized Facts for Modern Stock Markets}, 
  year={2023},
  volume={},
  number={},
  pages={1814-1823},
  doi={10.1109/BigData59044.2023.10386957}
}

@misc{Stylized_BitCoin1,
      title={Stylized Facts of High-Frequency Bitcoin Time Series}, 
      author={Yaoyue Tang and Karina Arias-Calluari and Michael S. Harré and Fernando Alonso-Marroquin},
      year={2024},
      eprint={2402.11930},
      archivePrefix={arXiv},
      primaryClass={q-fin.ST},
      url={https://arxiv.org/abs/2402.11930}, 
}

@article{Bouchaud_2001,
   title={Leverage Effect in Financial Markets: The Retarded Volatility Model},
   volume={87},
   ISSN={1079-7114},
   url={http://dx.doi.org/10.1103/PhysRevLett.87.228701},
   DOI={10.1103/physrevlett.87.228701},
   number={22},
   journal={Physical Review Letters},
   publisher={American Physical Society (APS)},
   author={Bouchaud, Jean-Philippe and Matacz, Andrew and Potters, Marc},
   year={2001},
   month=nov}

@article{Bouchaud_stylized,
title = {More stylized facts of financial markets: leverage effect and downside correlations},
author = {Jean-Philippe Bouchaud and M. Potters},
journal = {Physica A: Statistical Mechanics and its Applications},
volume = {299},
number = {1},
pages = {60-70},
year = {2001},
note = {Application of Physics in Economic Modelling},
issn = {0378-4371},
doi = {https://doi.org/10.1016/S0378-4371(01)00282-5},
url = {https://www.sciencedirect.com/science/article/pii/S0378437101002825},
}

@article{Bouchaud_universality_EV,
   title={Universality classes for extreme-value statistics},
   volume={30},
   ISSN={1361-6447},
   url={http://dx.doi.org/10.1088/0305-4470/30/23/004},
   DOI={10.1088/0305-4470/30/23/004},
   number={23},
   journal={Journal of Physics A: Mathematical and General},
   publisher={IOP Publishing},
   author={Bouchaud, Jean-Philippe and Mézard, Marc},
   year={1997},
   month=dec, pages={7997–8015} }

@article{Chicheportiche_2014,
   title={The fine-structure of volatility feedback I: Multi-scale self-reflexivity},
   volume={410},
   ISSN={0378-4371},
   url={http://dx.doi.org/10.1016/j.physa.2014.05.007},
   DOI={10.1016/j.physa.2014.05.007},
   journal={Physica A: Statistical Mechanics and its Applications},
   publisher={Elsevier BV},
   author={Chicheportiche, Rémy and Bouchaud, Jean-Philippe},
   year={2014},
   month=sep, pages={174–195} }

@article{Muller1997,
title = {Volatilities of different time resolutions — Analyzing the dynamics of market components},
journal = {Journal of Empirical Finance},
volume = {4},
number = {2},
pages = {213-239},
year = {1997},
note = {High Frequency Data in Finance, Part 1},
issn = {0927-5398},
doi = {https://doi.org/10.1016/S0927-5398(97)00007-8},
url = {https://www.sciencedirect.com/science/article/pii/S0927539897000078},
author = {Ulrich A. Müller and Michel M. Dacorogna and Rakhal D. Davé and Richard B. Olsen and Olivier V. Pictet and Jacob E. {von Weizsäcker}},
}

@article{Zumbach_2001,
   title={Heterogeneous volatility cascade in financial markets},
   volume={298},
   ISSN={0378-4371},
   url={http://dx.doi.org/10.1016/S0378-4371(01)00249-7},
   DOI={10.1016/s0378-4371(01)00249-7},
   number={3–4},
   journal={Physica A: Statistical Mechanics and its Applications},
   publisher={Elsevier BV},
   author={Zumbach, Gilles and Lynch, Paul},
   year={2001},
   month=sep, pages={521–529}
}

@article{Arneodo,
   title={'Direct' causal cascade in the stock market from the 'infrared' to the 'ultraviolet'},
   volume={2},
   ISSN={1434-6028},
   url={http://dx.doi.org/10.1007/s100510050250},
   DOI={10.1007/s100510050250},
   number={2},
   journal={The European Physical Journal B},
   publisher={Springer Science and Business Media LLC},
   author={Arnéodo, A. and Muzy, J.-F. and Sornette, D.},
   year={1998},

   month=may, pages={277–282} }

@article{Dacorogan_1993,
title = {A geographical model for the daily and weekly seasonal volatility in the foreign exchange market},
journal = {Journal of International Money and Finance},
volume = {12},
number = {4},
pages = {413-438},
year = {1993},
issn = {0261-5606},
doi = {https://doi.org/10.1016/0261-5606(93)90004-U},
url = {https://www.sciencedirect.com/science/article/pii/026156069390004U},
author = {Michael M. Dacorogna and Ulrich A. Müller and Robert J. Nagler and Richard B. Olsen and Olivier V. Pictet},
}

@article{ZHANG2019598,
title = {Stylised facts for high-frequency cryptocurrency data},
journal = {Physica A: Statistical Mechanics and its Applications},
volume = {513},
pages = {598-612},
year = {2019},
issn = {0378-4371},
doi = {https://doi.org/10.1016/j.physa.2018.09.042},
url = {https://www.sciencedirect.com/science/article/pii/S0378437118311567},
author = {Yuanyuan Zhang and Stephen Chan and Jeffrey Chu and Saralees Nadarajah}
}

@INPROCEEDINGS {Bjorn94,
author = {V. Bjorn},
booktitle = {Proceedings of 1995 Conference on Computational Intelligence for Financial Engineering (CIFEr)},
title = {Multiresolution methods for financial time series prediction},
year = {1995},
volume = {},
issn = {},
pages = {97},
doi = {10.1109/CIFER.1995.495258},
url={https://doi.ieeecomputersociety.org/10.1109/CIFER.1995.495258},
publisher = {IEEE Computer Society},
address = {Los Alamitos, CA, USA},
}

@ARTICLE{2020SciPy-NMeth,
  author  = {Virtanen, Pauli and Gommers, Ralf and Oliphant, Travis E. and
            Haberland, Matt and Reddy, Tyler and Cournapeau, David and
            Burovski, Evgeni and Peterson, Pearu and Weckesser, Warren and
            Bright, Jonathan and {van der Walt}, St{\'e}fan J. and
            Brett, Matthew and Wilson, Joshua and Millman, K. Jarrod and
            Mayorov, Nikolay and Nelson, Andrew R. J. and Jones, Eric and
            Kern, Robert and Larson, Eric and Carey, C J and
            Polat, {\.I}lhan and Feng, Yu and Moore, Eric W. and
            {VanderPlas}, Jake and Laxalde, Denis and Perktold, Josef and
            Cimrman, Robert and Henriksen, Ian and Quintero, E. A. and
            Harris, Charles R. and Archibald, Anne M. and
            Ribeiro, Ant{\^o}nio H. and Pedregosa, Fabian and
            {van Mulbregt}, Paul and {SciPy 1.0 Contributors}},
  title   = {{{SciPy} 1.0: Fundamental Algorithms for Scientific
            Computing in Python}},
  journal = {Nature Methods},
  year    = {2020},
  volume  = {17},
  pages   = {261--272},
  adsurl  = {https://rdcu.be/b08Wh},
  doi     = {10.1038/s41592-019-0686-2},
}

@Article{harris2020array,
 title         = {Array programming with {NumPy}},
 author        = {Charles R. Harris and K. Jarrod Millman and St{\'{e}}fan J.
                 van der Walt and Ralf Gommers and Pauli Virtanen and David
                 Cournapeau and Eric Wieser and Julian Taylor and Sebastian
                 Berg and Nathaniel J. Smith and Robert Kern and Matti Picus
                 and Stephan Hoyer and Marten H. van Kerkwijk and Matthew
                 Brett and Allan Haldane and Jaime Fern{\'{a}}ndez del
                 R{\'{i}}o and Mark Wiebe and Pearu Peterson and Pierre
                 G{\'{e}}rard-Marchant and Kevin Sheppard and Tyler Reddy and
                 Warren Weckesser and Hameer Abbasi and Christoph Gohlke and
                 Travis E. Oliphant},
 year          = {2020},
 month         = sep,
 journal       = {Nature},
 volume        = {585},
 number        = {7825},
 pages         = {357--362},
 doi           = {10.1038/s41586-020-2649-2},
 publisher     = {Springer Science and Business Media {LLC}},
 url           = {https://doi.org/10.1038/s41586-020-2649-2}
}

@inproceedings{mckinney2010data, 
  title={Data structures for statistical computing in python}, 
  author={McKinney, Wes}, 
  booktitle={Proceedings of the 9th Python in Science Conference}, 
  volume={445}, 
  pages={51--56}, 
  year={2010}, 
  organization={Austin, TX} 
}

@inproceedings{seabold2010statsmodels, 
  title={statsmodels: Econometric and statistical modeling with python}, 
  author={Seabold, Skipper and Perktold, Josef}, 
  booktitle={9th Python in Science Conference}, 
  year={2010}, 
}

@Article{Hunter:2007,
  Author    = {Hunter, J. D.},
  Title     = {Matplotlib: A 2D graphics environment},
  Journal   = {Computing in Science \& Engineering},
  Volume    = {9},
  Number    = {3},
  Pages     = {90--95},
  publisher = {IEEE COMPUTER SOC},
  doi       = {10.1109/MCSE.2007.55},
  year      = 2007
}

@book{python, 
 author = {Van Rossum, Guido and Drake, Fred L.}, 
 title = {Python 3 Reference Manual}, 
 year = {2009}, 
 isbn = {1441412697}, 
 publisher = {CreateSpace}, 
 address = {Scotts Valley, CA} 
}

@book{tsAnalysis_BrockwellDavis,
author = {Brockwell, Peter J and Davis, Richard A}, title = {Time series: theory and methods}, year = {1986}, isbn = {0387964061}, publisher = {Springer-Verlag}, address = {Berlin, Heidelberg} }

@article{Liu97,
   title={Correlations in economic time series},
   volume={245},
   ISSN={0378-4371},
   url={http://dx.doi.org/10.1016/S0378-4371(97)00368-3},
   DOI={10.1016/s0378-4371(97)00368-3},
   number={3–4},
   journal={Physica A: Statistical Mechanics and its Applications},
   publisher={Elsevier BV},
   author={Liu, Yanhui and Cizeau, Pierre and Meyer, Martin and Peng, C.-K. and Eugene Stanley, H.},
   year={1997},
   month={11}, 
   pages={437–440}
}

@article{abm_doyne,
  title={Agent-based modeling in economics and finance: Past, present, and future},
  author={Axtell, Robert L and Farmer, J Doyne},
  journal={Journal of Economic Literature},
  pages={1--101},
  year={2022},
  publisher={American Economic Association}
}

@article{Parkinson_estimator,
 ISSN = {00219398, 15375374},
 URL = {http://www.jstor.org/stable/2352357},
 author = {Michael Parkinson},
 journal = {The Journal of Business},
 number = {1},
 pages = {61--65},
 publisher = {University of Chicago Press},
 title = {The Extreme Value Method for Estimating the Variance of the Rate of Return},
 urldate = {2024-08-14},
 volume = {53},
 year = {1980}
}

@article{RogersSatchell,
 ISSN = {10505164},
 URL = {http://www.jstor.org/stable/2959703},
 author = {L. C. G. Rogers and S. E. Satchell},
 journal = {The Annals of Applied Probability},
 number = {4},
 pages = {504--512},
 publisher = {Institute of Mathematical Statistics},
 title = {Estimating Variance From High, Low and Closing Prices},
 urldate = {2024-08-14},
 volume = {1},
 year = {1991}
}

@article{intermittency,
author = {Bhansali, Rajendra and Holland, Mark and Kokoszka, Piotr},
year = {2007},
month = {01},
pages = {},
title = {Intermittency, Long-Memory and Financial Returns},
isbn = {978-3-540-22694-9},
journal = {Long Memory in Economics},
doi = {10.1007/978-3-540-34625-8_2}
}

@article{intermittency_turbulence,
author = {Pomeau, Yves and Manneville, Paul},
year = {1980},
month = {06},
pages = {},
title = {Intermittent transition to turbulence in dissipative dynamical systems. Commun. Math. Phys. 74, 189-197},
volume = {74},
journal = {Communications in Mathematical Physics},
doi = {10.1007/BF01197757}
}

@article{info_flow,
author = {Ghashghaie, Shoaleh and Breymann, Henriette and Peinke, Joachim and Talkner, Peter and Dodge, Yadolah},
year = {1996},
month = {06},
pages = {767-770},
title = {Turbulent Cascades in Foreign Exchange Markets},
volume = {381},
journal = {Nature},
doi = {10.1038/381767a0}
}

@misc{comments,
  TITLE = {{Comment on ``Turbulent cascades in foreign exchange markets'' [Ghashghaie et al., Nature 381, 767 (1996)]}},
  AUTHOR = {Arn{\'e}odo, A. and Bouchaud, J.-P. and Cont, R. and Muzy, J.-F. and Potters, M. and Sornette, D.},
  URL = {https://hal.science/hal-00126631},
  NOTE = {Commentaire, Reply mis sur le Web (arXiv)},
  HAL_LOCAL_REFERENCE = {SPEC-S96/153},
  YEAR = {1996},
  HAL_ID = {hal-00126631},
  HAL_VERSION = {v1},
}

@article{working1934,
author = {Holbrook Working},
title = {A Random-Difference Series for Use in the Analysis of Time Series},
journal = {Journal of the American Statistical Association},
volume = {29},
number = {185},
pages = {11--24},
year = {1934},
publisher = {ASA Website},
doi = {10.1080/01621459.1934.10502683},
URL = { 
        https://www.tandfonline.com/doi/abs/10.1080/01621459.1934.10502683
},
eprint = { 
        https://www.tandfonline.com/doi/pdf/10.1080/01621459.1934.10502683
}
}

@article{kendall1953,
author = {Kendall, M. G.},
title = {The Analysis of Economic Time-Series—Part I: Prices},
journal = {Journal of the Royal Statistical Society: Series A (General)},
volume = {116},
number = {1},
pages = {11-25},
doi = {https://doi.org/10.2307/2980947},
url = {https://rss.onlinelibrary.wiley.com/doi/abs/10.2307/2980947},
eprint = {https://rss.onlinelibrary.wiley.com/doi/pdf/10.2307/2980947},
year = {1953}
}

@ARTICLE{GoodhartFig,
title = {Every minute counts in financial markets},
author = {Goodhart, C. A. E. and Figliuoli, L.},
year = {1991},
journal = {Journal of International Money and Finance},
volume = {10},
number = {1},
pages = {23-52},
url = {https://EconPapers.repec.org/RePEc:eee:jimfin:v:10:y:1991:i:1:p:23-52}
}

@ARTICLE{andersen_boll1997,
title = {Intraday periodicity and volatility persistence in financial markets},
author = {Andersen, Torben and Bollerslev, Tim},
year = {1997},
journal = {Journal of Empirical Finance},
volume = {4},
number = {2-3},
pages = {115-158},
url = {https://EconPapers.repec.org/RePEc:eee:empfin:v:4:y:1997:i:2-3:p:115-158}
}

@incollection{Gency2001,
title = {Dedication},
editor = {Michel M. Dacorogna and Ramazan Gençay and Ulrich A. Müller and Richard B. Olsen and Olivier V. Pictet},
booktitle = {An Introduction to High-Frequency Finance},
publisher = {Academic Press},
address = {San Diego},
pages = {v},
year = {2001},
isbn = {978-0-12-279671-5},
doi = {https://doi.org/10.1016/B978-0-12-279671-5.50020-4},
url ={https://www.sciencedirect.com/science/article/pii/B9780122796715500204}
}

@ARTICLE{zhou1996,
title = {High-Frequency Data and Volatility in Foreign-Exchange Rates},
author = {Zhou, Bin},
year = {1996},
journal = {Journal of Business \& Economic Statistics},
volume = {14},
number = {1},
pages = {45-52},
url = {https://EconPapers.repec.org/RePEc:bes:jnlbes:v:14:y:1996:i:1:p:45-52}
}

@article{Heston1993,
  author = {Heston, S. L.},
  journal = {Review of Financial Studies},
  pages = {327--343},
  title = {A closed-form solution for options with stochastic
                 volatility with applications to bond and currency
                 options},
  volume = 6,
  year = 1993
}

@article{mandelbrot1997multifractal,
  title={A multifractal model of asset returns},
  author={Mandelbrot, Benoit B and Fisher, Adlai J and Calvet, Laurent E},
  year={1997},
  publisher={Cowles Foundation discussion paper}
}

@article{Grahovac_2016,
   title={Intermittency of Superpositions of Ornstein–Uhlenbeck Type Processes},
   volume={165},
   ISSN={1572-9613},
   url={http://dx.doi.org/10.1007/s10955-016-1616-7},
   DOI={10.1007/s10955-016-1616-7},
   number={2},
   journal={Journal of Statistical Physics},
   publisher={Springer Science and Business Media LLC},
   author={Grahovac, Danijel and Leonenko, Nikolai N. and Sikorskii, Alla and Tešnjak, Irena},
   year={2016},
   month=sep, pages={390–408} }

@article{fernando2019,
  title={Enriching Financial Datasets with Generative Adversarial Networks},
  author={de Meer Pardo, Fernando},
  year={2019}
}

@article{JurgOst_review,
  title={Generative adversarial networks in finance: an overview},
  author={Eckerli, Florian and Osterrieder, Joerg},
  journal={arXiv preprint arXiv:2106.06364},
  year={2021}
}

@article{JurgOst_crypt,
  title={The statistics of bitcoin and cryptocurrencies},
  author={Osterrieder, Joerg},
  journal={Available at SSRN 2872158},
  year={2016}
}

@misc{sherkar23,
      title={Study of Stylized Facts in Stock Market Data}, 
      author={Vaibhav Sherkar and Rituparna Sen},
      year={2023},
      eprint={2310.00753},
      archivePrefix={arXiv},
      primaryClass={q-fin.ST},
      url={https://arxiv.org/abs/2310.00753}, 
}

@book{PerBak_book,
    author = "Bak, Per",
    title = "{How nature works}",
    doi = "10.1007/978-1-4757-5426-1",
    year = "1996"
}

@book{zipf,
  address = {Oxford, England},
  author = {Zipf, George Kingsley},
  publisher = {Houghton Mifflin},
  series = {The psycho-biology of language: an introduction to dynamic philology},
  shorttitle = {The psycho-biology of language},
  title = {The psycho-biology of language : an introduction to dynamic philology},
  year = 1935
}

@ARTICLE{mandelbrot66,
title = {Forecasts of Future Prices, Unbiased Markets, and "Martingale" Models},
author = {Mandelbrot, Benoît},
year = {1965},
journal = {The Journal of Business},
volume = {39},
url = {https://EconPapers.repec.org/RePEc:ucp:jnlbus:v:39:y:1965:p:242}
}

@article{JJ_scipy_opt,
       author = {{Mor{\'e}}, Jorge J.},
        title = "{The Levenberg-Marquardt algorithm: Implementation and theory}",
    booktitle = {Lecture Notes in Mathematics, Berlin Springer Verlag},
         year = 1978,
       volume = {630},
        pages = {105-116},
          doi = {10.1007/BFb0067700},
       adsurl = {https://ui.adsabs.harvard.edu/abs/1978LNM...630..105M}
    ,journal = {Watson G.A. Ed. Numerical Analysis Springer Berlin 105-11}
}

@article{Dette2022effect,
address = {Berlin, Heidelberg},
author = {Holger Dette and Vasyl Golosnoy and Janosch Kellermann},
copyright = {http://www.econstor.eu/dspace/Nutzungsbedingungen; https://creativecommons.org/licenses/by/4.0/},
doi = {10.1007/s00184-022-00875-0},
issn = {1435-926X},
journal = {Metrika},
language = {eng},
number = {3},
pages = {315-342},
publisher = {Springer},
title = {The effect of intraday periodicity on realized volatility measures},
url = {https://hdl.handle.net/10419/307917},
volume = {86},
year = {2022}
}

@article{GARCH,
title = {Generalized autoregressive conditional heteroskedasticity},
journal = {Journal of Econometrics},
volume = {31},
number = {3},
pages = {307-327},
year = {1986},
issn = {0304-4076},
doi = {https://doi.org/10.1016/0304-4076(86)90063-1},
url = {https://www.sciencedirect.com/science/article/pii/0304407686900631},
author = {Tim Bollerslev}
}

@article{ARCH,
 ISSN = {00129682, 14680262},
 URL = {http://www.jstor.org/stable/1912773},
 author = {Robert F. Engle},
 journal = {Econometrica},
 number = {4},
 pages = {987--1007},
 publisher = {[Wiley, Econometric Society]},
 title = {Autoregressive Conditional Heteroscedasticity with Estimates of the Variance of United Kingdom Inflation},
 urldate = {2025-03-31},
 volume = {50},
 year = {1982}
}

@article{adf,
author = {Dickey, D. and Fuller, Wayne},
year = {1979},
month = {06},
pages = {},
title = {Distribution of the Estimators for Autoregressive Time Series With a Unit Root},
volume = {74},
journal = {JASA. Journal of the American Statistical Association},
doi = {10.2307/2286348}
}
\end{document}